\documentstyle[prb,preprint,eqsecnum,aps]{revtex}
\tighten
\begin{document}
\draft
\title{The role of Cooperons in the disordered electron problem:
logarithmic corrections to scaling near the metal-insulator transition}
\author{T.R.Kirkpatrick}
\address{Institute for Physical Science and Technology,\\
University of Maryland,\\
College Park, MD 20742}
\author{D.Belitz}
\address{Department of Physics and Materials Science Institute,\\
University ofOregon,\\
Eugene, OR 97403}

\date{\today}
\maketitle

\begin{abstract}
The effect of Cooperons on metal-insulator transitions (MIT) in disordered
interacting electronic systems is studied. It is argued that previous results
which concluded that Cooperons are qualitatively unimportant near the MIT
might not be correct, and that the problem is much more complicated than had
previously been
realized. Although we do not completely solve the Cooperon problem,
we propose a new approach that is at least internally consistent. Within this
approach we find that in all universality classes where Cooperons are present,
i.e. in the absence of magnetic impurities and magnetic fields, there are
logarithmic corrections to scaling at the MIT. This result is used for a
possible resolution of the so-called exponent puzzle for the conductivity
near the MIT. A discussion of the relationship between theory and
experiment is given. We also make a number of predictions concerning crossover
effects which occur when a magnetic field is applied to a system with
Cooperons.
\end{abstract}
\pacs{PACS numbers:}
\narrowtext
\section{INTRODUCTION}
\label{sec:I}

The current theoretical description of the metal-insulator transition (MIT)
problem asserts that the presence or absence of the particle-particle or
Cooper channel does not qualitatively modify the MIT.\cite{1,2} At first
sight this may seem surprising, as much of the early work in the modern
(post-1979) theory of the localization problem concentrated on the Cooper
channel and the backscattering or weak localization effects it produces.
\cite{3} Also, numerous experiments confirmed the presence of these weak
localization effects in weakly disordered metallic systems in the absence of
magnetic impurities and
magnetic fields.\cite{4} However, the assertion appears less surprising if
one recalls that electron-electron interaction effects in the presence of
disorder lead to many of the same effects as Cooperons.\cite{5}
If this is so in
the weak disorder regime, and if one acknowledges that electron-electron
interactions are in general relevant for the MIT, then it is
conceivable that Cooperons do not lead to any additional effects at the MIT
over and above those produced by the interplay of interactions and disorder
alone. This was in fact the conclusion reached by
Finkel'stein
\cite{1} and others \cite{2} who have argued that the Cooper channel is
irrelevant, in the sense of the renormalization group (RG), for the MIT,
that interaction effects effectively replace Cooperon effects near the MIT,
and that MIT with or without Cooperons are qualitatively the same.

In this paper we argue that the latter conclusion is probably not correct.
We present a consistent description within which the Cooper
propagator, or the effective Cooper interaction
amplitude, $\Gamma_c$, is a marginal operator rather than an irrelevant
one. This marginal operator leads to logarithmic corrections to scaling
that are characteristic
for those universality classes where Cooperons are present.
\par
Technically, we first show that the MIT problem in the presence of Cooperons
is substantially more complicated than had previously been realized. In
particular, the renormalization procedures used in existing treatments of the
problem do not lead to a finite renormalized field theory. We further argue
that even if this problem is ignored, the RG fixed point structure obtained in
previous treatments is not stable with respect to the consideration of higher
order terms that were neglected. We then present a partial solution to this
problem. We first point out that a principal unanswered question is that of
how many renormalization constant are needed to make the theory finite.
This is closely related to the physical question of whether $\Gamma_c$ is a
simple scaling variable near the MIT or whether it consists of several scaling
parts. Since at present we do not have a firm answer to this question, we
first derive RG flow equations for all other coupling constants which
appear in the theory. These flow equations can be expressed in terms of
$\Gamma_c$ whose scaling behavior is {\it a priori} unknown.
We then derive an Eliashberg-type integral
equation for $\Gamma_c$. General arguments lead us to the
conclusion that $\Gamma_c$ consists of several scaling parts, and that it
approaches its asymptotic value at the MIT, $\Gamma_c^*$, logarithmically
slowly.
\par
The most important conclusion from our considerations is that in systems
without magnetic impurities or magnetic fields there are logarithmic
corrections to scaling at the MIT. This in turn implies that it is virtually
impossible to experimentally reach the asymptotic critical scaling regime
for these universality classes, and that the existing experiments measure
effective exponents, rather than asymptotic critical exponents. We show that
these logarithmic corrections to scaling provide a possible resolution of
a long-standing problem. The critical exponent s for the conductivity in
Si:P \cite{6} and some other systems,
\cite{7,8} all of which are believed to be in
universality classes that contain Cooperons, is experimentally observed to be
smaller than 2/3, in apparent violation of a rigorous bound that requires
$s\ge 2/3$ in three-dimensional (3-D) systems.\cite{9}
We will see that logarithmic corrections to scaling can easily account
for an apparent or effective exponent $s_{eff}\approx 0.5$ even if the
true asymptotic critical exponent obeys $s\ge 2/3$.
The same conclusion was reached in a previous short
account of part of the work presented in this paper.\cite{10}

The plan of this paper is as follows. In Sec.\ \ref{sec:II} we
use Finkel'stein's effective field
theory for the MIT \cite{1} to calculate the perturbative corrections to the
coupling constants which appear in the theory to one-loop order.
In Sec.\ \ref{sec:III} we
use a normalization point RG approach to renormalize the field theory and to
derive RG flow equations for the coupling constants
in terms of the Cooper propagator $\Gamma_{c}$. We also
show how previous results can be obtained from certain assumptions and
approximations for $\Gamma_c$.
We find that these assumptions do not lead to a finite
renormalized theory, and that terms neglected in the previous treatments
invalidate this approach in any case. In Sec.\
\ref{sec:IV} we classify and discuss the
possible solutions of the integral equation for $\Gamma_c$. We argue that
generic solutions of the integral equation yield a $\Gamma_c$ that
approaches its fixed point value logarithmically slowly, which leads to
logarithmic corrections to scaling. In Sec.\
\ref{sec:V} we discuss the experimental
consequences of this result. We compare the theory with existing experiments
and suggest a number of new measurements to further test the theory. In
particular we discuss crossover
effects due to an external magnetic field which changes the
universality class and eliminates the logarithmic corrections to scaling. We
conclude in Sec.\
\ref{sec:VI} with a general discussion of our results as well as the
current status of the MIT problem.

\section{THE FIELD THEORY, AND THE LOOP EXPANSION}
\label{sec:II}

In the first part of this section we recall the basic field theoretic
description of the disordered electron problem
\cite{11,1} and derive the Gaussian
propagators of the field theory. We then explain how to obtain perturbative
corrections to the coupling constants by considering the generating
functional to one-loop order.

\subsection{The Model}
\label{subsec:II.A}

The existing theoretical description of the MIT is based on the usual
assumption of the theory of continuous phase transitions, viz. that the
physics near the transition is dominated by the low-lying excitations of the
system. As for the description of other phase transitions, the problem is then
reduced to obtaining a solution for an effective field theory for these slow
modes. Once the field theory has been identified, the technical
apparatus used to
obtain a solution is the RG. The initial problem is to obtain the field
theory, which requires a physical identification of the relevant slow modes.

In the field theory which we will use,
\cite{1,2} the slow modes in the metallic
phase are assumed to be the diffusion of mass, spin, and energy density. The
effective field theory for long-wavelength and low-frequency excitations that
describe how these diffusive processes change across an MIT is defined by an
action,
\begin{eqnarray}
S[Q]=&&-{1\over{2G}}\int d{\bf x}\;tr{\Bigl(}\nabla Q({\bf x}){\Bigr)}^2
     + 2H\int d{\bf x}\;tr{\Bigl(}\Omega Q({\bf x}){\Bigr)}
\nonumber\\
     &&-{\pi T \over 4}\sum_{u=s,t,c} K_{u} \int d{\bf x}\
     \bigl[ Q({\bf x}) \gamma^{(u)}Q({\bf x})].
\label{eq:2.1}
\end{eqnarray}
Here the field variable Q is an infinite matrix whose matrix elements,
$Q_{nm}^{\alpha \beta}$, are complex 4-by-4 matrices (spin-quaternions) which
comprise the spin and particle-hole degrees of freedom. The labels
$\alpha,\beta=1,2,\cdots,N$ denote replica labels. In deriving
Eq.\ (\ref{eq:2.1}), quenched
disorder has been integrated out by means of the replica trick,
\cite{12} and the limit
$N\rightarrow\infty$ is implied at the end of all calculations.
$n,m=-\infty,\cdots,+\infty$ are Matsubara frequency labels.
$\Omega={\bf I}\omega_n$ with
{\bf I} the identity matrix, $\omega_n=2\pi T(n+1/2)$, is a
fermionic frequency matrix,
and tr denotes a trace over all discrete degrees of freedom.
$[Q\gamma^{(u)}Q]$ in Eq.\ (\ref{eq:2.1}) is defined as,
\begin{mathletters}
\label{eqs:2.2}
\begin{eqnarray}
[Q\gamma^{(s)}Q]=\sum_{n_{1} n_{2} n_{3} n_{4}}%
     \delta_{n_{1}+n_{3},n_{2}+n_{4}}\sum_{\alpha}\sum_{r=0,3}(-1)^r
\nonumber\\
\times tr{\Bigl(}(\tau_{r} \otimes s_{0}) Q_{n_{1} n_{2}}^{\alpha \alpha}%
     {\Bigr)} tr{\Bigl(}(\tau_r \otimes s_0) Q_{n_{3} n_{4}}^{\alpha%
     \alpha}{\Bigr)}\quad,
\label{eq:2.2a}
\end{eqnarray}
\begin{eqnarray}
[Q&&\gamma^{(t)}Q]=-\sum_{n_{1} n_{2} n_{3} n_{4}}%
     \delta_{n_{1}+n_{3},n_{2}+n_{4}}\sum_{\alpha}%
     \sum_{r=0,3}(-1)^{r}
\nonumber\\
     &&\times \sum_{i=1}^{3}
     tr{\Bigl(}(\tau_{r}\otimes s_{i})Q_{n_{1} n_{2}}^{\alpha \alpha}
     {\Bigr)}
     tr{\Bigl(}(\tau_r \otimes s_i) Q_{n_{3} n_{4}}^{\alpha
     \alpha}{\Bigr)},
\label{eq:2.2b}
\end{eqnarray}
\begin{eqnarray}
[Q\gamma^{(c)}Q]=-\sum_{n_{1} n_{2} n_{3} n_{4}}
     \delta_{n_{1}+n_{2},n_{3}+n_{4}}\sum_{\alpha}\sum_{r=1,2}
\nonumber\\
\times tr_{s}{\Bigl[}tr_{\tau}{\Bigl(}(\tau_{r}\otimes s_{0})Q_{n_{1} n_{2}}
     ^{\alpha \alpha}{\Bigr)}
     tr_{\tau}{\Bigl(}(\tau_{r} \otimes s_{0})
      Q_{n_{3} n_{4}}^{\alpha \alpha}{\Bigr)}{\Bigr]},
\label{eq:2.2c}
\end{eqnarray}
\end{mathletters}
with $tr=tr_s tr_{\tau}$ where $tr_{\tau}$ acts only
on the $\tau$'s and $tr_s$
acts only on the s's. In Eqs.\ (\ref{eqs:2.2}), $\tau_0=s_0=\sigma_0$ and
$\tau_j=-s_j=-i\sigma_j\quad(j=1,2,3)$, with $\sigma_j$ the Pauli matrices.
The theory contains five coupling constants. $G=8/\pi\sigma_B$ with $\sigma_B$
the bare or self-consistent Born conductivity
is a measure of the disorder, and $H=\pi N_F/4$ plays the
role of a frequency coupling parameter with $N_F$ the bare density of states
(DOS) at the Fermi level. $K_s$ and $K_t$ are singlet and
triplet particle-hole interaction
constants, respectively, and $K_c$ is the singlet particle-particle or Cooper
channel interaction constant. At zero frequency the
triplet coupling constant in the particle-particle channel
vanishes due to the Pauli principle. A disorder generated, frequency dependent
triplet particle-particle interaction constant has been discussed elsewhere.
\cite{13} For simplicity we formulate the theory with a short-range
model interaction, i.e. the $K_{s,t,c}$ are simply numbers. For the more
realistic case of a Coulomb interaction $K_{s}$ is {\bf x}-dependent and
must be kept under the integral in
Eq.\ (\ref{eq:2.1}). Most results for this case are
easily obtained after all calculations have been performed by essentially
putting $K_{s}=-H$, \cite{1}
but occasionally subtle complications arise. Since
these are purely technical in nature, and decoupled from the Cooper channel
induced problems which are the subject of this paper we will not discuss
them. We will, however, give results for the Coulomb interaction case
in Sec.\ \ref{sec:III}.

The matrix Q is subject to the nonlinear constraints,
\begin{mathletters}
\label{eqs:2.3}
\begin{equation}
Q^{2}=1\quad,
\label{eq:2.3a}
\end{equation}
\begin{equation}
tr Q=0\quad,
\label{eq:2.3b}
\end{equation}
and satisfies,
\begin{equation}
Q^{\dag}=C^{T}Q^{T}C=Q\quad,
\label{eq:2.3c}
\end{equation}
with,
\begin{equation}
C=i(\tau_{1}\otimes s_{2})\quad.
\label{eq:2.3d}
\end{equation}
\end{mathletters}
The $\tau_{i}$ are the quaternion basis and span the particle-hole and
particle-particle space, while the $s_{i}$ serve as our basis in spin space.
For convenience we expand $Q_{nm}^{\alpha \beta}$ in this basis,
\begin{equation}
Q_{nm}^{\alpha \beta}=\sum_{r=0}^{3}\sum_{i=0}^{3}\;{^{i}_{r}Q_{nm}%
     ^{\alpha \beta}} (\tau_{r}\otimes s_{i})\quad.
\label{eq:2.4}
\end{equation}
{}From the first equality in Eq.\ (\ref{eq:2.3c}) it follows that the elements
of Q describing the particle-hole degrees of freedom are real, while those
describing the particle-particle degrees of freedom are purely imaginary,
\begin{mathletters}
\label{eqs:2.5}
\begin{equation}
^{i}_{r}Q_{nm}^{\alpha \beta *}=\;{^{i}_{r}Q_{nm}^{\alpha \beta}}%
     \qquad,\qquad(r=0,3)\quad,
\label{eq:2.5a}
\end{equation}
\begin{equation}
^{i}_{r}Q_{nm}^{\alpha \beta *}=-^{i}_{r}Q_{nm}^{\alpha \beta}%
     \qquad,\qquad(r=1,2)\quad,
\label{eq:2.5b}
\end{equation}
\end{mathletters}
In addition, from the hermiticity condition one obtains,
\begin{mathletters}
\label{eqs:2.6}
\begin{equation}
^{0}_{r}Q_{nm}^{\alpha \beta}={(-1)^{r}}\;{^{0}_{r}Q%
     _{mn}^{\beta \alpha}}\qquad,\qquad(r=0,3)\quad,
\label{eq:2.6a}
\end{equation}
\begin{equation}
^{i}_{r}Q_{nm}^{\alpha \beta}={(-1)^{r+1}}\;{^{i}_{r}Q%
     _{mn}^{\beta \alpha}}\quad,\quad(r=0,3\,;\;i=1,2,3)\quad,
\label{eq:2.6b}
\end{equation}
\begin{equation}
^{0}_{r}Q_{nm}^{\alpha \beta}={^{0}_{r}Q%
     _{mn}^{\beta \alpha}}\qquad,\qquad(r=1,2)\quad,
\label{eq:2.6c}
\end{equation}
\begin{equation}
^{i}_{r}Q_{nm}^{\alpha \beta}=-{^{i}_{r}Q%
     _{mn}^{\beta \alpha}}\qquad,\qquad(r=1,2\,;\;i=1,2,3)\quad.
\label{eq:2.6d}
\end{equation}
\end{mathletters}
The constraints given by Eqs.\ (\ref{eq:2.3a},\ref{eq:2.3b}) and the
hermiticity requirement, Eq.\ (\ref{eq:2.3c}),
can be eliminated by parametrizing the matrix Q by,\cite{14}
\begin{mathletters}
\label{eqs:2.7}
\begin{equation}
Q=\cases{(1-qq^{\dag})^{1/2} -1& for $n\ge 0$,\quad $m\ge 0$\cr%
     \qquad q& for $n\ge 0$,\quad $m<0$\cr%
     \qquad q^{\dag}& for $n<0$,\quad $m \ge 0$\cr%
     -(1-q^{\dag}q)^{1/2} -1& for $n<0$,\quad $m<0$\cr}%
     \qquad,
\label{eq:2.7a}
\end{equation}
Here the q are matrices with spin-quaternion valued elements $q_{nm}^{\alpha%
\beta}$;
$n=0,1,\cdots$; $m=-1,-2,\cdots$. Like the matrix Q, they can be expanded as
\begin{equation}
q^{\alpha \beta}_{nm}=\sum_{r=0}^{3}{\sum_{i=0}^{3}}\;%
     {^{i}_{r}q_{nm}^{\alpha \beta}}(\tau_{r}\otimes s_{i})%
     \quad.
\label{eq:2.7b}
\end{equation}
\end{mathletters}
Note that the q do not satisfy the symmetry relations given by
Eqs.\ (\ref{eqs:2.6}) for the Q.

We have so far given the theory for the so-called generic (G) universality
class which is realized by systems without magnetic fields, magnetic
impurities, or spin-orbit scattering. Apart from class G, the second
universality class with Cooperons is the one with strong spin-orbit
scattering (class SO). For class SO the action is shown as above, except
that the particle-hole spin triplet channel is absent, i.e. the
sum over the spin index i in Eqs.\ (\ref{eq:2.4}) and
(\ref{eq:2.7b}) is restricted
to $i=0$.\cite{15} In what follows we will give results for both class G and
class SO.

With the help of Eqs.\ (\ref{eqs:2.7}) one can expand the action in powers
of q,
\begin{equation}
S[Q]=\sum_{n=2}^{\infty}S_{n}[q]\quad,
\label{eq:2.8}
\end{equation}
where $S_{n}[q]\sim q^{n}$. We first concentrate on the Gaussian part of the
action,\cite{16}
\begin{mathletters}
\label{eqs:2.9}
\begin{equation}
S_{2}[q]=-4\int_{\bf p}\sum_{r,i}\sum_{1,2,3,4}%
     {^{i}_{r}q_{12}}({\bf p})\,{^{i}_{r}M_{12,34}}\,{^{i}_{r}q_{34}}%
     (-{\bf p})\quad.
\label{eq:2.9a}
\end{equation}
Here $\int_{\bf p}\equiv \int\,d{\bf p}/(2\pi)^{D}$, and
$1\equiv (n_{1},\alpha_{1})$, etc. The matrix {\bf M} is given by,
\widetext
\begin{equation}
^{i}_{0,3}M_{12,34}(p)={\delta_{1-2,3-4}\over G}%
     \bigl\{\delta_{13}[p^{2}+GH(\omega_{n_{1}}-\omega_{n_{2}})]%
     +\delta_{\alpha_{1} \alpha_{2}}\delta_{\alpha_{1}%
     \alpha_{3}}2\pi TGK_{\nu_{i}}\bigr\}\quad,
\label{eq:2.9b}
\end{equation}
where $\nu_{0}=s,\nu_{1,2,3}=t$, and
\begin{equation}
^{i}_{1,2}M_{12,34}(p)=-{\delta_{1+2,3+4}\over G}%
     \bigl\{\delta_{13}[p^{2}+GH(\omega_{n_{1}}-\omega_{n_{2}})]%
     +\delta_{\alpha_{1} \alpha_{2}}\delta_{\alpha_{1}%
     \alpha_{3}}\delta_{i,0}2\pi TGK_{c}\bigr\}\quad,
\label{eq:2.9c}
\end{equation}
\end{mathletters}
The Gaussian propagators are given by,
\widetext
\begin{mathletters}
\label{eqs:2.10}
\begin{eqnarray}
\Bigl<{^{i}_{r}q_{12}}({\bf p}_{1})^{j}_{s}q_{34}({\bf p}_{2})\Bigr>^{(2)}%
     =\int D[q]\;^{i}_{r}q_{12}({\bf p}_{1})\;^{j}_{s}q_{34}({\bf p}_{2})\;%
     e^{S_{2}[q]}\bigg/\int D[q]\;e^{S_{2}[q]}
\nonumber\\
     ={1 \over 8}\delta_{rs}\delta_{ij}%
     (2\pi)^{D}\delta({\bf p}_{1}+{\bf p}_{2})%
     \;{^{i}_{r}M^{-1}_{12,34}}(p_{1})\quad,
\label{eq:2.10a}
\end{eqnarray}
with,
\begin{equation}
^{i}_{0,3}M^{-1}_{12,34}(p)=\delta_{1-2,3-4}G\biggl(\delta%
     _{13}{\cal D}_{n_{1}-n_{2}}(p)+%
     {\delta_{\alpha_{1}\alpha_{2}}\over{n_{1}-n_{2}}}%
     \Delta{\cal D}^{\nu_{i}}_{n_{1}-n_{2}}(p)\biggr)\quad,
\label{eq:2.10b}
\end{equation}
\begin{eqnarray}
^{i}_{1,2}M^{-1}_{12,34}(p)=-\delta_{1+2,3+4}G\biggl(\delta%
     _{13}{\cal D}_{n_{1}-n_{2}}(p)&&
     -\delta_{\alpha_{1}\alpha_{2}}\delta_{i0}%
     {\cal D}_{n_{1}-n_{2}}(p){\cal D}_{n_{3}-n_{4}}(p)
\nonumber\\
     &&\times {G2\pi TK_{c}\over 1+G2\pi TK_{c}f_{n_{1}+n_{2}}(p)}%
     \biggr)\quad,
\label{eq:2.10c}
\end{eqnarray}
where,
\begin{equation}
f_{n}(p)=\sum_{n_{1} \ge 0, n_{2}<0}\delta_{n,n_{1}+n_{2}}{\cal D}
     _{n_{1}-n_{2}}(p)\quad.
\label{eq:2.10d}
\end{equation}
Here we have introduced the propagators,
\begin{equation}
{\cal D}_{n}(p)=[p^{2}+GH\Omega_{n}]^{-1}\quad,
\label{eq:2.10e}
\end{equation}
\begin{equation}
{\cal D}^{s,t}_{n}(p)=[p^{2}+G(H+K_{s,t})\Omega_{n}]^{-1}
     \quad,
\label{eq:2.10f}
\end{equation}
\begin{equation}
\Delta {\cal D}^{s,t}_{n}(p)={\cal D}_{n}^{s,t}(p)-{\cal D}_{n}(p)%
     \quad,
\label{eq:2.10g}
\end{equation}
\end{mathletters}
with $\Omega_{n}=2\pi Tn$ a bosonic Matsubara frequency.
Physically, ${\cal D}_{n}$, ${\cal D}_{n}^{s}$ and ${\cal D}_{n}^{t}$
are the energy, mass, and spin diffusion propagators.
\cite{17} Equations\ (\ref{eq:2.10b},\ref{eq:2.10c})
can be put into a more standard form by summing over, e.g., $n_{3}$
and $n_{4}$,
\begin{mathletters}
\label{eqs:2.11}
\begin{equation}
\sum_{n_{3}\ge 0,n_{4}<0}{^{i}_{0,3}M^{-1}_{12,34}}(p)/G=%
     (1-\delta_{\alpha_{1} \alpha_{2}})%
     {\cal D}_{n_{1}-n_{2}}(p) + \delta_{\alpha_{1} \alpha_{2}}%
     {\cal D}^{\nu_{i}}_{n_{1}-n_{2}}(p)%
     \quad,
\label{eq:2.11a}
\end{equation}
\begin{eqnarray}
\sum_{n_{3}\ge 0,n_{4}<0}{^{i}_{1,2}M^{-1}_{12,34}}(p)/G=-%
     (1-\delta_{\alpha_{1} \alpha_{2}})%
     {\cal D}_{n_{1}-n_{2}}(p)&& - \delta_{\alpha_{1} \alpha_{2}}%
     (1-\delta_{i0}){\cal D}_{n_{1}-n_{2}}(p)
\nonumber\\
     &&-\delta_{i0}\delta_{\alpha_{1} \alpha_{2}}%
     {{\cal D}_{n_{1}-n_{2}}(p)\over 1+G2\pi TK_{c}f_{n_{1}+n_{2}}(p)}
     \quad.
\label{eq:2.11b}
\end{eqnarray}
\end{mathletters}
\narrowtext
Examining the various terms in Eqs.\
(\ref{eqs:2.11}) we see that all of them have a
standard propagator structure except for the last contribution
in Eq.\ (\ref{eq:2.11b}).
In interpreting these propagators as having a standard structure, the
Matsubara frequencies in
Eqs.\ (\ref{eq:2.10e})-(\ref{eq:2.10g}) are taken to be analogous to a
magnetic field at a magnetic phase transition, i.e., the MIT occurs at
$\Omega_{n}\rightarrow 0$ and $\Omega_{n}$ or the temperature is a relevant
perturbation in the RG sense. Using
Eq.\ (\ref{eq:2.10a}), changing the sum to an
integral, and placing an ultraviolet cutoff, $\Omega_{0}$, on the resulting
frequency integrals shows that $2\pi Tf_{n_{1}+n_{2}}(p)$ in
Eq.\ (\ref{eq:2.11b})
diverges logarithmically in the long wavelength, low temperature limit. With
$K_{c}>0$, we see that the last term in
Eq.\ (\ref{eq:2.11b}) is logarithmically
small compared to the other terms in Eqs.\ (\ref{eqs:2.11}).

We conclude this subsection with two remarks. Firstly, the logarithm discussed
above that appears in the particle-particle
density correlation function is just the
usual BCS logarithm. However, since we consider a system with a repulsive
Cooper channel interaction, $K_{c}>0$, which is not superconducting in the
clean limit, this does not lead to a Cooper instability. Rather, the last
term in Eq.\ (\ref{eq:2.11b}) vanishes logarithmically in the limit
$p,T\rightarrow 0$.
If the structure of this term persists for disorder values up to the MIT,
and if it couples to the physical quantities like, e.g., the conductivity,
then it will lead to logarithmic corrections to scaling. Secondly, considering
the Gaussian theory one can already anticipate a fundamental problem with any
RG treatment of the field theory. To see this, note that at the Gaussian
level the two-point vertex functions are given by $S_{2}[q]$. Examining the
Eqs.\ (\ref{eqs:2.9})
we see that at this order no singularities, neither in the
ultraviolet nor in the infrared, are present in the vertex functions, and
there is no explicit cutoff dependence. This should be contrasted with the
corresponding two-point Gaussian propagators given by
Eqs.\ (\ref{eqs:2.10}). Because of
the last term in
Eq.\ (\ref{eq:2.10c}), both an infrared singularity and a dependence
on an ultraviolet cutoff appear. In the usual RG approach such cutoff
dependences are eliminated from the field theory by the introduction of
suitable renormalization constants.
Here, unusual features are that vertex functions and propagators
behave differently with respect to their cutoff dependence, and that
the cutoff dependent term is logarithmically small rather than large. In
previous RG treatments of this
problem, the procedure used was effectively to renormalize $K_{c}$ in Eqs.\
(\ref{eq:2.9c}) and (\ref{eq:2.10c})
such that the renormalized two-point {\it propagators} were
finite as $\Omega_{0}\rightarrow \infty$.
It is easy to see that such a procedure
leads to renormalized two-point {\it vertex functions} that contain a
singularity in this limit, cf.
Eq.\ (\ref{eqs:3.10}) below. In Sec.\ \ref{sec:III} we will further
discuss this problem, and we will propose an alternative renormalization
procedure.

\subsection{One-loop perturbation theory}
\label{subsec:II.B}

We will not be able to present a final solution of the Cooperon
renormalization problem. we will therefore
consider several renormalization procedures, and will discuss which ones
are at least internally consistent and which ones are not. One of these
procedures is based on perturbation theory for the generating functional for
the vertex functions, $\Gamma[q]$, which is closely related to the
thermodynamic potential. The perturbation expansion for $\Gamma[q]$ can be
generated by standard techniques.\cite{18} First, a source term with an
external potential $J_{nm}^{\alpha \beta}({\bf x})$ is added to the action so
that the average of $q_{nm}^{\alpha \beta}$, which we denote by
${\bar q}_{nm}^{\alpha \beta}$, is nonzero. We can then obtain a loop or
disorder expansion for $\Gamma[\bar q]$. It is convenient to expand
$\Gamma[\bar q]$ in powers of $\bar q$,
\widetext
\begin{equation}
\Gamma[{\bar q}]=\sum_{N=1}^{\infty}{1\over N!}%
     \sum_{1,2,\ldots,2N}\int d{\bf x}_{1}\cdots d{\bf x}_{N}\
     \Gamma^{(N)}_{1,2;3,4;\ldots;2N-1,2N}
     ({\bf x}_{1},\cdots,{\bf x}_{N})\
     {\bar q}_{12}({\bf x}_{1})\cdots {\bar q}_{2N-1,2N}({\bf x}_{N})
     \quad,
\label{2.12}
\end{equation}
where ${\Gamma}^{(N)}$ is the N-point vertex function.

To zero-loop order, one has
\begin{equation}
{\Gamma}_{0}[\bar q] = -S[\bar q]\quad.
\label{2.13}
\end{equation}
At higher-loop order the various coefficient in S acquire perturbative
corrections, and in addition structurally new terms are generated. For
simplicity we restrict our considerations to the two-point vertex function
${\Gamma}^{(2)}$, and to
the one-point vertex function $\Gamma^{(1)}$
which is related to the one-point propagator
$P^{(1)}=<^{0}_{0}Q_{nn}^{\alpha \alpha}({\bf x})>$.
At one-loop order ${\Gamma}^{(2)}$ is given by the
second derivative with respect to $\bar q$ of the
right-hand side of
Eq.\ (\ref{eq:2.9a}) with $q\rightarrow\bar q$, and with the matrix
M replaced by
\begin{mathletters}
\label{eqs:2.14}
\begin{equation}
^{i}_{0,3}M'_{12,34}(p)=\delta_{1-2,3-4}\biggl \{\delta_{13}\Bigl[%
     {p^{2} \over G_{n_{1}n_{2}}}+H_{n_{1}n_{2}}(\omega_{n_{1}}-%
     \omega_{n_{2}})\Bigr]
     + \delta_{\alpha_{1} \alpha_{2}}%
     \delta_{\alpha_{1} \alpha_{3}}2\pi TK^{\nu_{i}}_{n_{1}n_{2},
     n_{3}n_{4}}\biggr\}\quad,
\label{eq:2.14a}
\end{equation}
and,
\begin{equation}
^{i}_{1,2}M'_{12,34}(p)=-\delta_{1+2,3+4}\biggl \{\delta_{13}\Bigl[%
     {p^{2} \over G_{n_{1}n_{2}}}+H_{n_{1}n_{2}}(\omega_{n_{1}}-%
     \omega_{n_{2}})\Bigr]
     + \delta_{\alpha_{1} \alpha_{2}}%
     \delta_{\alpha_{1} \alpha_{3}}
     \delta_{i0}2\pi TK^{c}_{n_{1}n_{2},
     n_{3}n_{4}}\biggr\}\quad,
\label{eq:2.14b}
\end{equation}
\end{mathletters}
where $G_{n_{1}n_{2}}, H_{n_{1}n_{2}}$, and
$K^{s,t,c}_{n_{1}n_{2},n_{3}n_{4}}$ are
given by G, H, and $K_{s,t,c}$, respectively, plus frequency dependent
one-loop perturbative corrections. In our notation we have suppressed the
fact that these
corrections are in general also momentum dependent. In the absence of
Cooperons these corrections have been discussed in detail elsewhere.
\cite{16} In
general the momentum and frequency dependence of the corrections is quite
complicated. Here we just
give the results to leading order in $1/\epsilon$, with $\epsilon=D-2$. With
$\Lambda$ an ultraviolet momentum cutoff, and $\bar G =GS_{D}/(2\pi)^{D}$
with $S_{D}$ the surface of the D-dimensional unit sphere, we obtain,
\begin{mathletters}
\label{eqs:2.15}
\begin{eqnarray}
G_{n_{1}n_{2}}=G+G{\bar G \over 4 \epsilon} \Bigl [\Lambda^{\epsilon}
     &&- (GH\Omega_{n_{1}-n_{2}})^{\epsilon /2}\Bigr]
     -G^{2} \int_{\bf p} \bigl[I_{1}^{s}({\bf p},\Omega_{n_{1}-n_{2}})
     + 3I_{1}^{t}({\bf p},\Omega_{n_{1}-n_{2}})
\nonumber\\
     &&+I_{1}^{c}({\bf p},\Omega_{n_{1}-n_{2}})\bigr]
     + G^{2} \int_{\bf p} \bigl[I_{2}^{s}({\bf p},\Omega_{n_{1}-n_{2}})
     + 3I_{2}^{t}({\bf p},\Omega_{n_{1}-n_{2}})\bigr]\quad,
\label{eq:2.15a}
\end{eqnarray}
\begin{eqnarray}
H_{n_{1}n_{2}} = H + GH\int_{\bf p} \bigl[I_{1}^{s}({\bf p},
     \Omega_{n_{1}-n_{2}})&&
     + 3I_{1}^{t}({\bf p},\Omega_{n_{1}-n_{2}})
     + {1 \over 2}I_{1}^{c}({\bf p},\Omega_{n_{1}})
     +{1 \over 2}I_{1}^{c}({\bf p},\Omega_{-n_{2}})\bigr]
\nonumber\\
     &&- GH\int_{\bf p} \bigl[J_{2}^{s}({\bf p},\Omega_{n_{1}-n_{2}})
     + 3J_{2}^{t}({\bf p},\Omega_{n_{1}-n_{2}})\bigr]\quad,
\label{eq:2.15b}
\end{eqnarray}
\begin{equation}
K^{s}_{n_{1}n_{2},n_{3}n_{4}}=K_{s} + H - H_{n_{1}n_{2}}\quad,
\label{eq:2.15c}
\end{equation}
\begin{eqnarray}
K^{t}_{n_{1}n_{2},n_{3}n_{4}}=K_{t}&& +
     G(K_{t}-K_{s})\int_{\bf p}J_{1}({\bf p},\Omega_{n_{1}-n_{2}}) -
     {GK_{t} \over 2}\int_{\bf p}J_{3}({\bf p},\Omega_{n_{1}-n_{2}})
\nonumber\\
     &&-G2K_{t}\int_{\bf p}I_{1}^{c}({\bf p},\Omega_{n_{1}-n_{2}})%
     -{G \over 2}(H+K_{t})^{2}\int_{\bf p}J_{2}^{c}({\bf p},
     \Omega_{n_{1}-n_{2}})\quad,
\label{eq:2.15d}
\end{eqnarray}
\begin{equation}
K^{c}_{n_{1}n_{2},n_{3}n_{4}}=K_{c}-{\bar G \over 8\epsilon}
     \bigl[\Lambda^{\epsilon}-(GH\Omega_{n_{3}-n_{2}})^{\epsilon/2}
     \bigr](K_{s}-3K_{t}) + %
     G\int_{\bf p}J_{1}^{c}({\bf p},\Omega_{n_{3}-n_{2}})\quad,
\label{eq:2.15e}
\end{equation}
\end{mathletters}
Here, $\Omega_{0}=O(\Lambda^{2} /GH)$ is a cutoff frequency, and,
\begin{mathletters}
\label{eqs:2.16}
\begin{equation}
I_{1}^{s,t}({\bf p},\Omega_{n}) = {1 \over 8}%
     \sum_{m=n}^{\Omega_{0}/2 \pi T}{1 \over m}%
     \Delta {\cal D}_{m}^{s,t}({\bf p})\quad,
\label{eq:2.16a}
\end{equation}
\begin{equation}
I_{1}^{c}({\bf p},\Omega_{n}) = - {1 \over 8} G 2 \pi T%
     \sum_{m=n}^{\Omega_{0}/2\pi T}%
     {K_{c} \over 1+G2\pi TK_{c}f_{-m}({\bf p})}%
     \Bigl( {\cal D}_{m}({\bf p}) \Bigr)^{2}\quad,
\label{eq:2.16b}
\end{equation}
\begin{equation}
I_{2}^{s,t}({\bf p},\Omega_{n})=-{G^{2} \over 8}(K_{s,t})^{2}2\pi T
     \sum_{m=n}^{\Omega_{0}/2\pi T}%
     \Omega_{m} {\cal D}_{m}^{s,t}({\bf p})\Bigl(%
     {\cal D}_{n+m}({\bf p})\Bigr)^{2}
     \times \Bigl[1 - 2p^{2}{\cal D}_{n+m}({\bf%
     p})\Bigr]\quad,
\label{eq:2.16c}
\end{equation}
\begin{equation}
J_{2}^{s,t}({\bf p},\Omega_{n}) = {1 \over 8}(K_{s,t}/H)%
     {2\pi T \over \Omega_{n}}%
     \sum_{m=n}^{\Omega_{0}/2\pi T}%
     \Bigl[\Delta {\cal D}_{m}^{s,t}({\bf p}) + GK_{s,t}\Omega_{m}%
     {\cal D}_{m}^{s,t}({\bf p}){\cal D}_{m+n}({\bf p})\Bigr]%
     \quad,
\label{eq:2.16d}
\end{equation}
\begin{equation}
J_{1}({\bf p},\Omega_{n})={2\pi T \over 8\Omega_{n}}%
     \sum_{m=1}^{n-1}\Bigl[\bigl(1-{m \over n}\bigr){\cal D}_{m+n}({\bf p})
     +{m \over n}{\cal D}_{m}({\bf p})\Bigr]\quad,
\label{eq:2.16e}
\end{equation}
\begin{eqnarray}
J_{3}({\bf p}&&,\Omega_{n}) =- \sum_{m=n}^{\Omega_{0}/2\pi T}
     {1 \over m}\Delta {\cal D}_{m}^{t}({\bf p})
     -GK_{t}\sum_{m=n}^{\Omega_{0}/2\pi T}{\cal D}_{m}({\bf p})
     {\cal D}_{m+n}({\bf p})
     +{G \over 2}K_{s}\sum_{m=n}^{\Omega_{0}/2\pi T}
     {\cal D}_{m}^{s}({\bf p}){\cal D}_{m+n}({\bf p})
\nonumber\\
     &&+{G \over 4}K_{t}\sum_{m=n}^{\Omega_{0}/2\pi T}{n \over m}
     {\cal D}_{m}^{s}
     ({\bf p}){\cal D}_{m+n}({\bf p})
     +{G \over 4}K_{t}\sum_{m=n}^{\Omega_{0}/2\pi T}\Bigl(2-{m \over n}\Bigr)
     {\cal D}_{m}^{t}({\bf p})
     {\cal D}_{m+n}({\bf p})\quad,
\label{eq:2.16f}
\end{eqnarray}
\begin{equation}
J_{2}^{c}({\bf p},\Omega_{n}) = 2\pi T \sum_{m=n}^{\Omega_{0}/2\pi T}
     \Omega_{m}\Bigl({\cal D}_{m}({\bf p})\Bigr)^{3}%
     {K_{c} \over 1+K_{c}G2\pi Tf_{-m}({\bf p})}%
     \quad,
\label{2.16g}
\end{equation}
\begin{equation}
J_{1}^{c}({\bf p},\Omega_{n}) = {1 \over 2}G2\pi T
     \sum_{m=n}^%
     {\Omega_{0}/2\pi T}{1 \over m}\Delta {\cal D}_{m}^{s}({\bf p})%
     {K_{c} \over 1+G2\pi TK_{c}f_{-m}({\bf p})}\quad,
\label{eq:2.16h}
\end{equation}
\end{mathletters}
For the one-point vertex function one finds,
\begin{equation}
\Gamma^{(1)}(\Omega_{n}) = 1 +%
     G\int_{\bf p}[I_{1}^{s}({\bf p},\Omega_{n})
     +3I_{1}^{t}({\bf p},\Omega_{n}) + I_{1}^{c}({\bf p},\Omega_{n})]
     \quad.
\label{eq:2.17}
\end{equation}
\narrowtext
In giving Eqs.\ (\ref{eqs:2.15})-(\ref{eq:2.17}) we have neglected terms
that are finite in
$D=2$ as $(\Lambda,\Omega_{0}) \rightarrow \infty$,
and a delta function constraint (cf. Eqs.\ (\ref{eqs:2.14})) is understood
in Eqs.\ (\ref{eq:2.15c},\ref{eq:2.15d},\ref{eq:2.15e}). Some of these terms
depend on $n_{1}$ and $n_{2}$ separately, not just on the difference
$n_{1}-n_{2}$. In
Eq.\ (\ref{eq:2.15b}) we have written a separate dependence on
$n_{1}$ and $n_{2}$ explicitly for later reference. Also, the complete
frequency dependence of $K_{n_{1}n_{2},n_{3}n_{4}}^{t}$ and
$K_{n_{1}n_{2},n_{3}n_{4}}^{c}$ is more complicated than the one shown.
However, for most of our purposes it is sufficient to treat all 'external'
frequencies as equal, and we do not have to deal with the (substantial)
complications that arise from the full perturbation theory.

Finally, let us discuss one important point. To one-loop order the two-point
propagators are given by the right-hand side of Eq.\ (\ref{eq:2.10a}) with the
inverse matrix $M^{-1}$ replaced by the matrix ${M'}^{-1}$ which contains
the perturbative corrections. For the particle-hole degrees of freedom one
can show in general that, except for irrelevant terms, the matrix ${M'}^{-1}$
has the same form as the matrix $M^{-1}$, with the only difference being
that the corrected coupling constants appear in ${M'}^{-1}$. The underlying
reason for this feature is the conservation laws for mass, spin, and energy
density. For the particle-particle degrees of freedom the situation is
different. There are no conservation laws which guarantee
that the form of the last term in
Eq.\ (\ref{eq:2.10c}) will not change at higher orders in the loop expansion.
In general, the matrix $_{1,2}{(M')}^{-1}$ is given by
Eq.\ (\ref{eq:2.10c}) with the replacement,
\begin{mathletters}
\label{eqs:2.18}
\begin{equation}
{K_{c} \over 1+G2\pi TK_{c}f_{n_{1}+n_{2}}(p)}%
     \rightarrow \Gamma^{c}_{n_{1}n_{2},n_{3}n_{4}}(p)%
     \quad,
\label{eq:2.18a}
\end{equation}
where $\Gamma^{(c)}$ satisfies an Eliashberg-type integral equation,
\widetext
\begin{equation}
\Gamma^{c}_{n_{1}n_{2},n_{3}n_{4}} + 2\pi T%
     \sum_{n'_{1}\ge 0,n'_{2}<0}%
     \Gamma^{c}_{n_{1}n_{2},n'_{1}n'_{2}}\ %
     {K^{c}_{n'_{1}n'_{2},n_{3}n_{4}}%
     \delta_{n_{3}+n_{4},n'_{1}+n'_{2}}\over%
     p^{2}/G_{n'_{1}n'_{2}} + H_{n'_{1}n'_{2}}%
     (\omega_{n'_{1}}-\omega_{n'_{2}})}
     =K^{c}_{n_{1}n_{2},n_{3}n_{4}}\quad.
\label{eq:2.18b}
\end{equation}
\narrowtext
Here we have again suppressed the momentum dependence of $\Gamma^{c}$.
Note that if we make the substitutions
$K^{c}_{n'_{1}n'_{2},n_{3}n_{4}}\rightarrow K_{c}$,
$H_{n'_{1}n'_{2}}\rightarrow H$, $G_{n'_{1}n'_{2}}\rightarrow G$,
replace the remaining sum by an integral, let $p,T\rightarrow 0$, and use
an ultraviolet cutoff $\Omega_{0}$ on the frequency integral, then
$\Gamma^{c}$ has the standard BCS form,
\begin{equation}
\Gamma^{c}(p,\omega_{n_{1}+n_{2}})={K_{c}\over%
     1+{\gamma_{c}^{(0)}\over 2}\ln\Bigl({\Omega_{0}\over%
     D^{(0)}p^{2}+\mid \omega_{n_{1}+n_{2}}\mid}\Bigr)}\quad,
\label{eq:2.18c}
\end{equation}
\end{mathletters}
with $\gamma_{c}^{(0)}=K_{c}/H$, and $D^{(0)}=1/GH$.

The actual Cooper propagator to all orders in perturbation theory would have
the simple form given by
Eq.\ (\ref{eq:2.18c}) only if the coupling constants $K^{c}$
and H were constant to all orders. In general, the structure of the Cooper
propagator will be more complicated and to obtain it one has to solve the
integral equation, Eq.\ (\ref{eq:2.18b}). For later reference we symmetrize
the equation by defining
\begin{mathletters}
\label{eqs:2.19}
\begin{equation}
\gamma_{n_{1}n_{2},n_{3}n_{4}}={K^{c}_{n_{1}n_{2},n_{3}n_{4}}\over%
     \bigl[ H_{n_{1}n_{2}}H_{n_{3}n_{4}}\bigr]^{1/2}}\quad,
\label{eq:2.19a}
\end{equation}
and,
\begin{equation}
\tilde \Gamma_{n_{1}n_{2},n_{3}n_{4}}%
     ={\Gamma^{c}_{n_{1}n_{2},n_{3}n_{4}}\over%
     \bigl[ H_{n_{1}n_{2}}H_{n_{3}n_{4}}\bigr]^{1/2}}\quad.%
\label{eq:2.19b}
\end{equation}
\end{mathletters}
For $T\rightarrow 0$ the integral equation for $\tilde \Gamma$ then reads,
\begin{mathletters}
\label{eqs:2.20}
\begin{eqnarray}
\tilde \Gamma (\omega,\Omega,\omega'') +&& \int_{0}^{\Omega_{0}} d\omega'%
     {\gamma (\omega,\Omega,\omega')\tilde \Gamma (\omega',\Omega,\omega'')%
     \over D(\omega',-\Omega - \omega')p^{2} + 2\omega' + \Omega}%
\nonumber\\
     &&=\gamma(\omega,\Omega,\omega'')\quad,
\label{eq:2.20a}
\end{eqnarray}
with,
\begin{equation}
D(\omega',-\Omega - \omega')=[G(\omega',-\Omega - \omega')%
     H(\omega',-\Omega - \omega')]^{-1}\quad,
\label{eq:2.20b}
\end{equation}
Since we consider the zero temperature limit we have made the replacements,
\begin{eqnarray}
2\pi T(n_{1}+n_{2})=2\pi T(n_{3}+n_{4})\rightarrow -\Omega
\nonumber\\
2\pi Tn_{3}\rightarrow \omega
\nonumber\\
2\pi Tn'_{1}\rightarrow \omega'
\nonumber\\
2\pi Tn_{1}\rightarrow \omega''
\nonumber\\
\tilde \Gamma_{n_{1}n_{2},n_{3}n_{4}}\rightarrow \tilde \Gamma%
     (\omega,\Omega,\omega'')
\label{eq:2.20c}
\end{eqnarray}
\end{mathletters}
etc., and for definiteness we have assumed $\Omega\ge 0$.
We stress again that Eq.\ (\ref{eq:2.20a})
has the form of an Eliashberg equation with a repulsive kernel.

\section{THE RENORMALIZATION GROUP FLOW EQUATIONS}
\label{sec:III}

In the first part of this section we review the general procedure of the field
theoretic RG as applied to a nonlinear sigma-model. We then use a
normalization point RG procedure to obtain RG flow equations for all of the
parameters that appear in the field theory except for the Cooper propagator
$\tilde \Gamma$. Finally, we discuss the reasons why previous attempts to
derive a flow equation for $\tilde \Gamma$ are problematic.

\subsection{The Nonlinear Sigma-Model and the Renormalization Group}
\label{subsec:III.A}

The basic philosophy behind the field theoretic RG approach is to eliminate
all singular ultraviolet cutoff dependences, i.e. singularities of the
theory as $(\Lambda,\Omega_{0})\rightarrow \infty$, by introducing
renormalization constants $Z_{i},\;(i=1,2,\cdots)$. RG flow equations are
then obtained by examining how the $Z_{i}$ depend on the cutoff.\cite{18} In
principle this approach is equivalent to the Wilsonian RG, which examines
how the theory changes when the ultraviolet cutoff is changed from, e.g.,
$\Lambda$ to $\Lambda/b$ with b a RG rescaling factor.\cite{19} However, only
a limited amount of work has been done on the formal relationship between
these two formulations of the RG.\cite{20}

A point of fundamental importance in any RG approach to a field theory is
that one has to determine how many independent scaling operators there are.
In the field theoretic RG approach, one needs to know how many renormalization
constants are needed to make the theory finite. The field theory defined by
Eqs.\ (\ref{eq:2.1})-(\ref{eqs:2.3}) has the form of a nonlinear
sigma-model with perturbing
operators. Let us first consider the pure sigma-model part, i.e. the first
term on the r.h.s. of Eq.\
(\ref{eq:2.1}) with the constraints given by Eqs.\ (\ref{eqs:2.3}).
This term is invariant under the symplectic symmetry group Sp(8nN), with
N the number of replicas and 2n the number of Matsubara frequencies. This
model is well known to be renormalizable with two renormalization constants,
one for the coupling constant G (the disorder) and one for the
renormalization of the Q-field.\cite{21} The second term on the r.h.s. of
Eq.\ (\ref{eq:2.1})
breaks the symmetry to
\[
\underbrace{Sp(4N)\otimes Sp(4N)\otimes \cdots \otimes Sp(4N)}
     _{2n\ {\rm times}}\quad.
\]
This term does not require any additional renormalization constants because
the coupling constant H in
Eq.\ (\ref{eq:2.1}) just multiplies the basic Q-field, and
therefore the renormalization of H is determined by the field renormalization
constant.\cite{21,22} This model represents the noninteracting localization
problem.\cite{11} From a physical
point of view the model has a rather restrictive
property: The only interaction taken into account is the elastic
electron-impurity scattering, and consequently the different Matsubara
frequencies in
Eq.\ (\ref{eq:2.1}) are decoupled. Effectively, n is held fixed, and this
is crucial for the simple renormalization properties of the noninteracting
model. The situation remains relatively simple if further terms are added
which respect the non-mixing of the frequencies. In this case, one needs
one additional renormalization constant for each operator that
represents a different irreducible representation of Sp(8nN).\cite{21}

The situation changes fundamentally with the addition of the last term in
Eq.\ (\ref{eq:2.1}). Physically, this term describes the electron-electron
interaction and hence the exchange of energy between electrons. Technically,
this leads to a coupling between the Matsubara frequencies, and an
examination of the perturbation theory shows that this term introduces new
infrared and ultraviolet singularities as $n\rightarrow \infty$. In the
absence of interactions, singularities arise only from momentum integrations
and the symmetry group is fixed to be Sp(8nN). With interactions there are
singularities due to both momentum and frequency integrations, and the
symmetry properties of the model change continuously during the RG procedure.
As a consequence, the results mentioned above concerning the
renormalizability of nonlinear sigma-models with perturbing operators,
which apply to models with a fixed symmetry, are inapplicable. No general
results concerning the number of renormalization constants needed are
available, and it is unclear how to renormalize the model given by the full
Eq.\ (\ref{eq:2.1}). The symmetry arguments quoted above can provide only a
lower bound on the number of Z's needed to renormalize the theory, and it is
not known whether the model is renormalizable with a finite number of
renormalization constants.

All RG treatments of the field theory defined by
Eqs.\ (\ref{eq:2.1})-(\ref{eqs:2.3}) so far
have ignored this general renormalizability problem. They have assumed,
explicitly or implicitly, that the full model is still renormalizable
with one extra renormalization constant for each interaction coupling
constant which is added. In addition, H acquires a renormalization
constant of its own once interactions are present.
In the absence of Cooperons, i.e. in a theory
which contains only $K_{s}$ and $K_{t}$, there is empirical evidence based
on perturbation theory for this assumption being correct.\cite{14,23}
In the presence
of Cooperons things are more complicated as we will see. For pedagogical
reasons, and to make contact with previous work, we nevertheless proceed
for a while using this assumption. It is most convenient to use a
normalization point RG. The renormalized disorder, frequency coupling,
interaction constants, and q-fields are denoted by g, h, $k_{s}$, $k_{t}$,
$k_{c}$, and $q_{R}$, respectively. They are defined by,
\begin{mathletters}
\label{eqs:3.1}
\begin{equation}
{\bar G}=\mu^{-\epsilon} Z_{g} g\quad,
\label{eq:3.1a}
\end{equation}
\begin{equation}
H=Z_{h} h\quad,
\label{eq:3.1b}
\end{equation}
\begin{equation}
K_{s,t,c}=Z_{s,t,c} k_{s,t,c}\quad,
\label{eq:3.1c}
\end{equation}
\begin{equation}
q=Z^{1/2} q_{R}\quad,
\label{eq:3.1d}
\end{equation}
\end{mathletters}
where $\mu$ is an arbitrary momentum scale that is introduced in
Eq.\ (\ref{eq:3.1a})
to make g dimensionless. The renormalized N-point vertex functions are
related to the bare ones by,
\widetext
\begin{equation}
\Gamma^{(N)}_{R}(p,\Omega_{n};g,h,k_{s},k_{t},k_{c};\mu,\Lambda)%
     =Z^{N/2}\,\Gamma^{(N)}(p,\Omega_{n};G,H,K_{s},K_{t},K_{c};\Lambda)%
     \quad.
\label{eq:3.2}
\end{equation}
The theory is called renormalizable with these renormalization constants
if all of the $\Gamma^{(N)}_{R}$ are finite as $\Lambda \rightarrow \infty$
for fixed renormalized coupling constants. The Z's in
Eqs.\ (\ref{eqs:3.1}) are not
unique. We fix them by the following normalization conditions for the
two-point vertex functions,
\begin{mathletters}
\label{eqs:3.3}
\begin{equation}
{1 \over 8}{\partial \over \partial p^{2}}\;%
     {^{i}_{r}(\Gamma_{R}^{(2)})^{\alpha \beta
     ,\alpha \beta}_{n_{1}n_{2},n_{1}n_{2}}}(p)%
     {\mid}_{p=0,\;\omega_{n_{1}}-\omega_{n_{2}}%
     =\mu^{D}/gh,\;\omega_{n_{1}}+\omega_{n_{2}}=0}=%
     {\mu^{\epsilon} \over g}S_{D}/(2\pi)^{D}\quad,%
\label{eq:3.3a}
\end{equation}
\begin{equation}
{1 \over 8}{\partial \over \partial (\omega_{n_{1}}-\omega_{n_{2}})}\;%
     {^{i}_{r}(\Gamma_{R}^{(2)})^{\alpha \beta,%
     \alpha \beta}_{n_{1}n_{2},n_{1}n_{2}}}(p)%
     {\mid}_{\alpha \not= \beta,\;p=0,\;\omega_{n_{1}}-\omega_{n_{2}}%
     =\mu^{D}/gh,\;\omega_{n_{1}}+\omega_{n_{2}}=0}=h\quad,%
\label{eq:3.3b}
\end{equation}
\begin{equation}
{1 \over 8}{\Bigl[{^{i}_{0}(\Gamma_{R}^{(2)})^{\alpha \alpha, \alpha
     \alpha}
     _{n_{1}n_{2},n_{1}n_{2}}}(p=0)%
     -^{i}_{0}(\Gamma_{R}^{(2)})^{\alpha \beta,%
     \alpha \beta}_{n_{1}n_{2},n_{1}n_{2}}(p=0)%
     {\mid}_{\alpha \not= \beta}\Bigr]}_{\omega_{n_{1}}-\omega_{n_{2}}%
     =\mu^{D}/gh,\;\omega_{n_{1}}+\omega_{n_{2}}=0}
     =2\pi Tk_{\nu_{i}}\quad,
\label{eq:3.3c}
\end{equation}
\begin{equation}
{1 \over 8}{\Bigl[{^{0}_{1}(\Gamma_{R}^{(2)})^{\alpha \alpha, \alpha \alpha}
     _{n_{1}n_{2},n_{1}n_{2}}}(p=0)%
     - ^{0}_{1}(\Gamma_{R}^{(2)})^{\alpha \beta,%
     \alpha \beta}_{n_{1}n_{2},n_{1}n_{2}}(p=0)%
     {\mid}_{\alpha \not= \beta}\Bigr]}_{\omega_{n_{1}}+\omega_{n_{2}}%
     =-\mu^{D}/gh,\;\omega_{n_{1}}-\omega_{n_{2}}=0}
     =-2\pi Tk_{c}\ ,
\label{eq:3.3d}
\end{equation}
\narrowtext
These conditions determine the renormalization constants $Z_{g,h,s,t,c}$.
The wavefunction or field renormalization constant Z we fix by a
normalization condition for the one-point vertex function $\Gamma^{(1)}$.
We require,
\begin{equation}
^{0}_{0}(\Gamma_{R}^{(1)})^{\alpha \alpha}_{nn} {\mid}_{\Omega%
     _{n}=\mu^{D}/gh}=1\quad.
\label{eq:3.3e}
\end{equation}
\end{mathletters}
Note that the normalization conditions given by Eqs.\ (\ref{eqs:3.3})
are the usual
ones: At scale $\mu$ the renormalized vertex functions are taken to have
their tree level structure.

{}From Eqs.\ (\ref{eqs:2.14}), (\ref{eqs:2.15}), (\ref{eqs:3.1}),
(\ref{eq:3.2}), and (\ref{eqs:3.3}) the Z's can be determined.
We obtain,
\begin{mathletters}
\label{eqs:3.4}
\begin{equation}
Z = \Bigl(1/\Gamma^{(1)}(\Omega_{n}=\mu^{D}/gh)\Bigr)^{2}%
     \quad,
\label{eq:3.4a}
\end{equation}
\begin{eqnarray}
Z_{g}=Z - {\mu^{\epsilon} \over g}{S_{D} \over (2\pi)^{D}}[G_{n_{1}-n_{2}}-
     G]_{\omega_{n_{1}}-\omega_{n_{2}}=\mu^{D}/gh}%
\nonumber\\
     +O(g^{2})\quad,
\label{eq:3.4b}
\end{eqnarray}
\begin{equation}
Z_{h} = Z^{-1} - {1 \over h}[H_{n_{1}-n_{2}}-
     H]_{\omega_{n_{1}}-\omega_{n_{2}}=\mu^{D}/gh}%
     +O(g^{2})\quad,
\label{eq:3.4c}
\end{equation}
\begin{eqnarray}
Z_{s,t,c} = Z^{-1} - {1 \over k_{s,t,c}}[K_{n_{1}n_{2},n_{3}n_{4}}^{s,t,c}
     -&&K_{s,t,c}]_{\omega_{n_{1}}-\omega_{n_{2}}=\mu^{D}/gh}
\nonumber\\
     &&+O(g^{2})\quad,
\label{eq:3.4d}
\end{eqnarray}
\end{mathletters}
The fastest way to derive the RG flow equations is to switch from our
cutoff regularized theory to a minimal subtraction scheme. We can do so
by putting $\epsilon < 0$ in
Eqs.\ (\ref{eqs:3.4}), letting $\Lambda \rightarrow \infty$,
and then analytically continuing to $\epsilon > 0$. We find,
\begin{mathletters}
\label{eqs:3.5}
\begin{equation}
Z = 1 - {g \over 4\epsilon}\Bigl[{-2 \over \epsilon} + 3l_{t}%
     + \tilde \Gamma \Bigr]\quad,
\label{eq:3.5a}
\end{equation}
\begin{equation}
Z_{g}=1+{g \over 4\epsilon} \biggl\{ 5-3(1+1/\gamma_{t})l_{t}-%
     {\tilde \Gamma \over 2} \biggr\} + O(g^{2})\quad,
\label{eq:3.5b}
\end{equation}
\begin{equation}
Z_{h} = 1 + {g \over 8\epsilon}\Bigl[-1+3\gamma_{t}+\tilde \Gamma%
     \Bigr]\quad,
\label{eq:3.5c}
\end{equation}
\begin{equation}
Z_{s} = Z_{h}\quad,
\label{eq:3.5d}
\end{equation}
\begin{equation}
Z_{t} = 1 + {g \over 2\epsilon}
     \Bigl[{1 \over 4\gamma_{t}} + {1 \over 4} + \gamma_{t}
     +\tilde \Gamma \Bigl({1 \over 4\gamma_{t}} + 1 + {1 \over 2}\gamma_{t}
     \Bigr)\Bigr]\quad,
\label{eq:3.5e}
\end{equation}
\begin{eqnarray}
Z_{c} = 1 + {g \over 4\epsilon}\Bigl[{-2 \over \epsilon} + 3l_{t}
     + \tilde \Gamma \Bigr] + {g \over 4\epsilon}
     (1+3\gamma_{t})/\gamma_{c}
\nonumber\\
     + {g \over \epsilon^{2}}\tilde \Gamma /\gamma_{c}
     \quad,
\label{eq:3.5f}
\end{eqnarray}
\end{mathletters}
with $l_{t}=\ln(1+\gamma_{t})$, $\gamma_{t,c}=k_{t,c}/h$. In giving Eqs.\
(\ref{eqs:3.5}), and (\ref{eqs:3.6}) below, we have for simplicity
neglected terms of order $g\tilde\Gamma^{n}$ with $n\ge 2$. The omission
of these terms does not affect our conclusions.

The one-loop RG flow equations follow from
Eqs.\ (\ref{eqs:3.1}) and (\ref{eqs:3.5}) in the usual
way.\cite{18} With $b\sim \mu^{-1}$ and $x=\ln \,b$ one obtains,
\begin{mathletters}
\label{eqs:3.6}
\begin{equation}
{dg \over dx}=-\epsilon g + {g^{2}\over 4}\Bigl\{5-3(1+1/\gamma_{t})%
     l_{t}-{\tilde \Gamma \over 2}\Bigr\}+O(g^{3})\quad,
\label{eq:3.6a}
\end{equation}
\begin{equation}
{dh \over dx}={gh \over 8}[3\gamma_{t} -1+\tilde \Gamma] + O(g^{2})%
     \quad,
\label{eq:3.6b}
\end{equation}
\begin{equation}
{d\gamma_{t} \over dx}={g \over 8}(1+\gamma_{t})^{2}+{g \over 8}%
     (1+\gamma_{t})(1+2\gamma_{t})\tilde \Gamma +O(g^{2})\quad,
\label{eq:3.6c}
\end{equation}
\begin{eqnarray}
{d\gamma_{c} \over dx}={g \over 4}(3\gamma_{t}+1)+{g\gamma_{c} \over 8}%
     \Bigl\{-{4 \over \epsilon}+6l_{t}+1-3\gamma_{t}+\tilde \Gamma%
\nonumber\\
     +{8 \over \epsilon}{\tilde \Gamma \over \gamma_{c}}\Bigr\}+O(g^{2})%
     \quad.
\label{eq:3.6d}
\end{eqnarray}
\end{mathletters}
In Eqs.\ (\ref{eqs:3.5}), (\ref{eqs:3.6}) $\tilde \Gamma$ is the Cooper
propagator at scale $\mu$:
\begin{mathletters}
\label{eqs:3.7}
\begin{equation}
\tilde \Gamma \equiv%
     \tilde \Gamma (\mu)=\tilde \Gamma(\omega,\Omega,\omega') {\mid}_%
     {\omega=\omega'=\mu^{D}/2gh=\Omega/2}\quad.
\label{eq:3.7a}
\end{equation}
To lowest order in the disorder one has,
\begin{equation}
\tilde \Gamma (\mu)={\gamma_{c} \over 1+\gamma_{c} x}%
     +O(g)\quad.
\label{eq:3.7b}
\end{equation}
\end{mathletters}
In giving Eqs.\ (\ref{eqs:3.5}), (\ref{eqs:3.6}) we have
specialized to the case of a Coulomb interaction
between the electrons. In this case a compressibility sum rule enforces
$\gamma_{s}=k_{s}/h=-1$,\cite{1} and the terms $\ln(1+\gamma_{s})$
that appear in doing the integrals become
$-2/\epsilon$ in Eqs.\ (\ref{eqs:3.6}). The presence of these terms in the
RG flow
equation for $\gamma_{c}$ reflects the well-known $(\ln)^{2}$ singularity
that exists in the disorder expansion of the single-particle
DOS in D=2.\cite{5} It has sometimes been claimed that these DOS effects
are absent in all flow equations for the interactions constants.\cite{24}
We stress
that this statement depends on what quantity exactly one tries to
derive a flow equation for. We find
that it is true for $\gamma_{s}$ and $\gamma_{t}$, but not for the
Cooperon amplitude $\gamma_{c}$. It is also important to note that the
presence of these $1/\epsilon$-terms in the flow equations {\it per se}
does not create any problems. They do appear, e.g., in the renormalization
of the single-particle DOS, for which a careful application of the
RG \cite{25} leads to results that are consistent both internally and with
those obtained by other methods.\cite{1}

The Eqs.\ (\ref{eqs:3.6}) are valid for the universality class G. For the
spin-orbit class SO the analogous results are,
\begin{mathletters}
\label{eqs:3.8}
\begin{equation}
{dg \over dx}=-\epsilon g + {g^{2}\over 8}[1-\tilde \Gamma]%
     +O(g^{3})\quad,
\label{eq:3.8a}
\end{equation}
\begin{equation}
{dh \over dx}=-{gh \over 8}[1-\tilde \Gamma] + O(g^{2})%
     \quad,
\label{eq:3.8b}
\end{equation}
\begin{equation}
{d\gamma_{c} \over dx}={g \over 4}+{g\gamma_{c} \over 8}%
     \Bigl\{-{4 \over \epsilon}+1+\tilde \Gamma%
     +{8 \over \epsilon}{\tilde \Gamma \over \gamma_{c}}\Bigr\}%
     +O(g^{2})\quad.\eqnum{3.8d}
\label{eq:3.8d}
\end{equation}
\end{mathletters}
Note that the Eqs.\ (\ref{eqs:3.8}) for the universality class SO have a
fixed point describing a MIT,\cite{2} while the Eqs.\ (\ref{eqs:3.6})
for the generic universality class
do not. The description of the MIT in the class G is more complicated.\cite
{26} For our present purposes we do not need a detailed description of the MIT
and the Eqs.\ (\ref{eqs:3.6}) are sufficient.

\subsection{A Scaling equation for $\tilde \Gamma$}
\label{subsec:III.B}

The RG flow equations given by
Eqs.\ (\ref{eqs:3.6}) are not closed because they contain
the Cooper propagator $\tilde \Gamma$.
Ref.\ \onlinecite{10} just used Eq.\ (\ref{eq:3.7b}) in
Eqs.\ (\ref{eqs:3.6}). In general this is not satisfactory, since for a
consistent one-loop RG description one needs $d\tilde \Gamma /dx$ to
one-loop order. In Ref.\ \onlinecite{10} we argued that this was justified
near the MIT since we found that $\gamma_{c}$ approaches a finite FP
value at the transition. Here we try
to improve on this point. In principle, $\tilde \Gamma$ can be
determined by solving the integral equation given by
Eqs.\ (\ref{eqs:2.20}) together
with Eqs.\ (\ref{eqs:3.6}) and (\ref{eqs:3.7}).
This structure is very different from the one
usually encountered in RG approaches where the flow equations are an
autonomous set of coupled differential equations. In this subsection we
discuss attempts to reduce the integral equation for $\tilde \Gamma$ to a
single differential equation. We note that for generic integral equations
this can not be done. What has effectively been assumed in the previous
literature \cite{1,2} is that the reduction is possible in this case.
As we will see, this reduction leads to severe structural problems
which went unnoticed in the previous treatments quoted
since the techniques used were not sensitive to them. In the next section
we will therefore turn to a different method, which determines the behavior
of $\tilde \Gamma$ directly from
Eqs.\ (\ref{eqs:2.20}). This will qualitatively yield
the same result as inserting
Eq.\ (\ref{eq:3.7b}) into Eqs.\ (\ref{eqs:3.6}). Here we investigate
possible avenues for deriving a flow equation for $\tilde \Gamma$
strictly in order to make contact with previous work.

Within the normalization point RG approach we can formally obtain a
differential equation for $\tilde \Gamma$ by renormalizing the Cooper
propagator rather than the Cooper vertex function. We impose a normalization
condition,
\begin{mathletters}
\label{eqs:3.9}
\begin{equation}
\tilde \Gamma (\mu)=\gamma_{c}\quad,
\label{eq:3.9a}
\end{equation}
and define a renormalization constant $\bar Z_{c}$ by
\begin{equation}
\gamma_{c}^{(0)}=\bar Z_{c} \gamma_{c}\quad.
\label{eq:3.9b}
\end{equation}
\end{mathletters}
Notice that this approach {\it assumes} that $\tilde \Gamma$ is a simple
scaling quantity. To zeroth order in the disorder
Eqs.\ (\ref{eqs:3.7}) and (\ref{eqs:3.9}) give,
\begin{mathletters}
\label{eqs:3.10}
\begin{equation}
\bar Z_{c}=(1-\gamma_{c}x)^{-1}\quad,
\label{eq:3.10a}
\end{equation}
\begin{equation}
{d \gamma_{c} \over dx}=-\gamma_{c}^{2}\quad.
\label{eq:3.10b}
\end{equation}
\end{mathletters}
A few remarks are in order in the context of Eq.\ (\ref{eq:3.10b}): (1)
It has the standard form of a RG flow equation for a
marginal operator. (2) In this approach the RG is used to obtain the BCS
logarithm. This is in contrast to, e.g.,
Eq.\ (\ref{eq:2.11b}) where we obtained the
BCS logarithm at Gaussian order by inverting the vertex function. (3) In
the present subsection the physical meaning of $\gamma_{c}$ is
different from the rest of this paper. Due to
Eq.\ (\ref{eq:3.9a}) it plays the role
of the Cooper propagator rather
than that of the Cooper interaction amplitude that appeared in the previous
subsection. (4) A crucial question is what the structure of the higher
loop corrections to
Eq.\ (\ref{eq:3.10b}) will be. In a particular approximation
that has been made in the literature and which we will discuss below,
$\gamma_{c}$ flows to a nonzero
fixed point value, $\gamma_{c} \rightarrow \gamma_{c}^{*}$. In this case
Eq.\ (\ref{eq:3.10a}) naively implies that $\bar Z_{c}$
has a Cooper-type singularity at a finite scale $x=\ln b=1/\gamma_{c}^{*}$
which does not correspond to any physical phase transition.
Of course, this conclusion is in general not necessarily correct, since
terms of higher order in the disorder could change the behavior of
$\bar Z_{c}$. Nevertheless, we will see that
the appearance of such an unphysical singularity represents a serious problem.
\cite{footnote1} We will discuss this point in connection with
Eq.\ (\ref{eq:3.14}) below, and in the Conclusion.

Within this approach the one-loop RG flow equation
can be obtained by using
Eqs.\ (\ref{eqs:2.15}) and (\ref{eq:2.19a}) in Eq.\ (\ref{eq:2.20a}),
and iterating to first order in the disorder.
{}From Eqs.\ (\ref{eqs:3.1}) and
(\ref{eqs:3.9}) one can then obtain a flow equation for
$\gamma_{c}$. The resulting equation is quite complicated. Since we have
come to the conclusion that this is not a viable approach
we will not give a complete discussion, but rather illustrate only a few
points. First, let us make contact with previous work \cite{1,2} by retaining
only the first two terms in
Eq.\ (\ref{eq:2.15e}), neglecting the corrections to H,
putting $p=0$, and working in D=2. In this
approximation, $\gamma$ in
Eq.\ (\ref{eq:2.20a}) is given to leading logarithmic accuracy by,
\begin{mathletters}
\label{eqs:3.11}
\begin{equation}
\gamma(\omega,\Omega,\omega')=\gamma_{c}^{(0)}-a_{1}\ln \biggl({\omega %
     + \omega' + \Omega \over \Omega_{0}} \biggr)\quad,
\label{eq:3.11a}
\end{equation}
with,
\begin{equation}
a_{1}=\cases{(g/16) (-\gamma_{s}+3\gamma_{t})& for class G\cr%
     (g/16)& for class SO\cr}\quad,
\label{eq:3.11b}
\end{equation}
\end{mathletters}
for the generic and spin-orbit universality classes, respectively. Using
Eq.\ (\ref{eq:3.11a}) in the procedure discussed above, we obtain the
following one-loop flow equation for $\gamma_{c}$,
\begin{mathletters}
\label{eqs:3.12}
\begin{equation}
{d\gamma_{c} \over dx}=-\gamma_{c}^{2} [1-a_{1}c]-4a_{1}\gamma_{c} \ln 2 %
     +2a_{1} + O(g^{2},x^2e^{-2x})\ ,
\label{eq:3.12a}
\end{equation}
with,
\begin{equation}
c=\int_{0}^{\infty} {dz \over z+1/2} \ln \biggl({z+3/2 \over z+1/2} \biggr)%
     - {\pi^{2} \over 12} \quad.
\label{eq:3.12b}
\end{equation}
\end{mathletters}
Two things should be noted: (1) The coefficients in
Eq.\ (\ref{eq:3.12a}) are universal
and, consequently, this equation has the form of a standard RG flow
equation. (2) At the MIT in, e.g., the SO universality class, $a_{1}$, which
is of O(g), is a constant of order $g^{*}=O(\epsilon)$.
If Eq.\ (\ref{eq:3.12a}) is valid, then this implies that $\gamma_{c}$ goes to
a fixed point value atthe MIT which is of order $\epsilon^{1/2}$. In
a strict $\epsilon$-expansion the terms of $O(g\gamma_{c})$ and higher on the
r.h.s of Eq.\ (\ref{eq:3.12a}) can be neglected,
and to leading order the equation takes the form,
\begin{equation}
{d\gamma_{c} \over dx}=-\gamma_{c}^{2} + 2a_{1}\quad.
\eqnum{3.12a'}
\label{eq:3.12a'}
\end{equation}
This result is identical to those in
Refs.\ \onlinecite{1} and \onlinecite{2}, except for the prefactor
of the $a_{1}$, which is 4 in
Refs.\ \onlinecite{1,2}. This difference is due to the fact
that for a Coulomb interaction, which was considered in these references,
an additional term appears in
Eq.\ (\ref{eq:2.15e}). This leads to a more complicated
kernel than the one in Eq.\ (\ref{eq:3.11a}),
and to an additional factor of 2 in Eq.\ (\ref{eq:3.12a'}). Since this
difference is not relevant for our purposes we
restrict ourselves to the simpler kernel for the short-range case,
Eq.\ (\ref{eq:3.11a}).

In the approximation discussed above one finds $\gamma_{c} \rightarrow$ const.
at the MIT. As noted below
Eqs.\ (\ref{eqs:3.10}), this result creates some problems.
To illustrate this point we add to
Eq.\ (\ref{eq:3.11a}), via Eq.\ (\ref{eq:2.19a}), the
Cooper propagator contribution to $H_{n_{1}n_{2}}$ and $H_{n_{3}n_{4}}$,
i.e. the integral over $I_{1}^{c}$ in Eq.\ (\ref{eq:2.15b}).
The result is a kernel in Eq.\ (\ref{eq:2.20a}) which is given by,
\widetext
\begin{mathletters}
\label{eqs:3.13}
\begin{eqnarray}
\gamma(\omega,\Omega,\omega')=\gamma_{c}^{(0)}-a_{1}\ln \biggl( {\omega
     + \omega' + \Omega \over \Omega_{0}} \biggr)
     +a_{2}\gamma_{c}^{(0)} \Bigl(\ln [1&&-{1 \over 2}\gamma_{c}^{(0)} \ln (
     \omega/\Omega_{0})]
\nonumber\\
     &&+\ln [1-{1 \over 2}\gamma_{c}^{(0)} \ln (%
     \omega'/\Omega_{0})] \Bigr)\quad,
\label{eq:3.13a}
\end{eqnarray}
with,
\begin{equation}
a_{2}=g/16\quad.
\label{eq:3.13b}
\end{equation}
\end{mathletters}
The last two terms in
Eq.\ (\ref{eq:3.13a}) occur because of the dependence of $\gamma$
on the Cooper propagator. In the last term in
Eq.\ (\ref{eq:3.13a}) we have neglected
a dependence on $\Omega$ which is irrelevant for our purposes.
Using Eq.\ (\ref{eq:3.13a}) in the RG procedure yields the flow equation,
\begin{eqnarray}
{d\gamma_{c} \over dx}=-\gamma_{c}^{2}[1-a_{1}c-2a_{2}] - 4a_{1}\gamma_{c}%
     \ln 2 + 2a_{1}
     -a_{2}\gamma_{c}^{2} \int_{0}^{b^{2}}{dz \over (z+1/2)^{2}}%
     \ln [1-&&{\gamma_{c} \over 2} \ln(2z)]
\nonumber\\
     && + O(g^{2})\quad.
\label{eq:3.14}
\end{eqnarray}
\narrowtext
The crucial point is that as $x=\ln b \rightarrow \infty$ the last term in
Eq.\ (\ref{eq:3.14}) does not exist because of a singularity at
$b \sim exp(1/\gamma_{c}) \sim exp(1/\epsilon^{1/2})$. Notice that this
breakdown of the RG flow equation for the Cooper propagator is nonperturbative
in nature. Previous attempts to derive a RG flow equation for the Cooper
propagator \cite{1,2} were based on a frequency-momentum
shell RG approach. This method
is by necessity restricted to low order perturbative expansions in both
g and $\gamma_{c}$ and cannot be used to discuss the singularity in
Eq.\ (\ref{eq:3.14}).
Furthermore, if one expands the last term in Eq.\ (\ref{eq:3.14}) in powers
of $\gamma_{c}$ then a non-Borel summable divergent series is obtained.
The structure of this singularity resembles what is known as the renormalon
problem in quantum field theory.\cite{18} We also note that this singularity
is obviously related to the one discussed below Eq.\ (\ref{eq:3.10b}).

In the following section we will argue that the Cooper propagator actually
consists of multiple scaling parts, and that any theoretical approach that
does not acknowledge this feature is invalid.
It is unclear whether or not the singularity
discussed in connection with
Eq.\ (\ref{eq:3.14}) is related to this problem, which in
turn is related to the question of whether or not the theory is
renormalizable. Usually it is necessary to
go to higher than one-loop order in order to verify the presence of
multiple scaling parts. In the present case, however, the fact that a
renormalization of the propagator \cite{1,2} and the vertex function,
\cite{10} respectively, led to different results is already an indication of
their presence.

\section{THE ELIASHBERG EQUATION, AND LOGARITHMIC CORRECTIONS TO SCALING}
\label{sec:IV}

In the first part of this section we discuss some general features of the
Eliashberg equation with a repulsive kernel given by Eq.\ (\ref{eq:2.20a}).
We then give theoretical arguments in favor of the existence of
logarithmic corrections to scaling at the MIT, and discuss their
experimental relevance.

\subsection{The Eliashberg Equation}
\label{subsec:IV.A}

In order to complete the RG description started in Sec.\
\ref{subsec:III.A}, the scaling
properties of the Cooper propagator given by
Eqs.\ (\ref{eqs:2.19}), (\ref{eqs:2.20}), and (\ref{eqs:3.7})
need to be determined. In
Sec.\ref{subsec:III.B} we showed why previous attempts to
reduce the integral equation for $\tilde \Gamma$ to a differential
equation are not satisfactory. Here we pursue a different approach. We
acknowledge that one has to actually solve the integral equation in order
to obtain information about $\tilde \Gamma$. Since this is very difficult
to do in general, we classify possible solutions for different behaviors
of the kernel $\gamma(\omega,\Omega,\omega')$.

To simplify our considerations we work at zero momentum. We further
specialize to a model with a separable kernel. After drawing some
conclusions for this special case we will argue that these are actually
generic. The main points can be illustrated using a kernel that is a sum
of two separable parts,
\begin{mathletters}
\label{eqs:4.1}
\begin{equation}
\gamma(\omega,\Omega,\omega'')=f_{1}(\omega,\Omega)\,f_{1}(\omega'',\Omega)%
     +f_{2}(\omega,\Omega)\,f_{2}(\omega'',\Omega)\quad.
\label{eq:4.1a}
\end{equation}
Note that Eq.\ ({eq:4.1a}) satisfies the symmetry requirement,
\begin{equation}
\gamma(\omega,\Omega,\omega'')=\gamma(\omega'',\Omega,\omega)\quad.
\label{eq:4.1b}
\end{equation}
\end{mathletters}
Eq.\ (\ref{eq:4.1b}) is a consequence of the symmetry property
$K^{c}_{n_{1}n_{2},n_{3}n_{4}} = K^{c}_{n_{3}n_{4},n_{1}n_{2}}$,
which in term follows from
Eq.\ (\ref{eq:2.2c}). Inserting
Eq.\ (\ref{eq:4.1a}) into Eq.\ (\ref{eq:2.20a}) leads to a separable
integral equation that can be easily solved. We obtain,
\begin{mathletters}
\label{eqs:4.2}
\begin{eqnarray}
\tilde \Gamma (\omega,\Omega,\omega'') =
     \bigl[(1+J_{1})(1+J_{2}) - J_{3}^{2}\bigr]^{-1}
\nonumber\\
     \times \Bigl[ f_{1}(\omega,\Omega) f_{1}(\omega'',\Omega)\,(1+J_{2})
\nonumber\\
     + f_{2}(\omega,\Omega) f_{2}(\omega'',\Omega)\,(1+J_{1})
\nonumber\\
     - [f_{1}(\omega,\Omega)
     f_{2}(\omega'',\Omega) + f_{1}(\omega'',\Omega)f_{2}(\omega,\Omega)]%
     J_{3}\Bigr]\quad,
\label{eq:4.2a}
\end{eqnarray}
with,
\begin{equation}
J_{1,2}=\int_{0}^{\Omega_{0}}d\omega'\;{[f_{1,2}(\omega',\Omega)]^{2}%
     \over 2\omega' + \Omega}\quad,
\label{eq:4.2b}
\end{equation}
and,
\begin{equation}
J_{3}=\int_{0}^{\Omega_{0}}d\omega'\;{f_{1}(\omega',\Omega)f_{2}(%
     \omega',\Omega) \over 2\omega' + \Omega}\quad.
\label{eq:4.2c}
\end{equation}
\end{mathletters}

In the previous literature it has been suggested that $\tilde \Gamma (\mu)$
given by Eq.\ (\ref{eq:3.7a}) goes to a finite fixed point value at the MIT
(cf. the discussion in the previous subsection), and that
the approach to criticality is characterized by a conventional power law.
To see how this type of behavior can be realized from
Eqs.\ (\ref{eqs:4.1}), (\ref{eqs:4.2}), we
specialize to criticality, and put $f_{2}=0$. If $f_{1}$ diverges like, e.g.,
\begin{equation}
f_{1}(\omega,\Omega) \sim (\omega + \Omega)^{-\alpha}\qquad,%
     (\alpha >0)\quad,
\label{eq:4.3}
\end{equation}
then the fixed point (FP) value of $\tilde \Gamma$ is,
\begin{mathletters}
\label{eqs:4.4}
\begin{equation}
\tilde \Gamma^{*}=2^{1-2\alpha}3^{-2\alpha}%
     \Biggl[ \int_{0}^{\infty} {dx \over%
     (x+1/2)(x+1)^{2\alpha}}\Biggr]^{-1}\quad,
\label{eq:4.4a}
\end{equation}
and near the FP $\tilde\Gamma$ satisfies the flow equation,
\begin{equation}
{d\tilde \Gamma \over d\,\ln \mu^{2}}=\tilde \Gamma^{2} 2\alpha%
     /\tilde \Gamma^{*} - 2\alpha \tilde \Gamma\quad,
\label{eq:4.4b}
\end{equation}
\end{mathletters}
that is, $\tilde \Gamma - \tilde \Gamma^{*} \sim \mu^{4\alpha}$.
Similarly, if $f_{2}=0$ and $f_{1}$ vanishes at the MIT
(Eq.\ (\ref{eq:4.3}) with
$\alpha <0$), then $\tilde \Gamma$ also vanishes and satisfies a flow
equation with universal coefficients. The marginal, logarithmic approach
to zero occurs if $f_{2}$ vanishes and $f_{1}$ approaches a constant at the
MIT.

If we assume that these asymptotic results are not tied to the separable
kernel, but are generic properties of the general Eliashberg equation,
then we have the following situation:
If $\gamma$ diverges (vanishes) at the MIT, then $\tilde \Gamma$ has a finite
(zero) FP value, and if $\gamma$ goes to a constant then $\tilde \Gamma$
vanishes logarithmically slowly. In all of these cases $\tilde \Gamma$
satisfies an autonomous differential equation with universal coefficients.

However, from a more general point of view, going beyond the asymptotic
behavior, one expects a more complex result. For instance, even if
$\gamma \rightarrow \gamma^{*}$ at the MIT one expects a correction that
vanishes either as a power law or as a logarithm. In fact, the
Eq.\ (\ref{eq:3.6d})
predicts this kind of behavior. In our model calculation this happens if
$f{_2} \not= 0$. With
Eqs.\ (\ref{eqs:4.2}) it is easy to see that $\tilde \Gamma (\mu)$
does not satisfy an autonomous differential equation if $f_{2} \not= 0$.
Of course, this just reflects the obvious fact that in general an integral
equation cannot be reduced to a single differential equation.

We further note that even if $\gamma$ diverges at the MIT, then one
generically still expects a finite subleading contribution. Using this in
either
Eq.\ (\ref{eq:4.2a}) or in the actual Eliashberg equation,
Eq.\ (\ref{eq:2.20a}), one
finds that (1) $\tilde \Gamma (\mu)$ does not satisfy a single differential
equation, and (2) $\tilde \Gamma$ approaches a finite FP value
$\tilde \Gamma^{*}$, but logarithmically slowly so.

We conclude that for both the case where $\gamma$ diverges and where it
approaches a constant at the MIT one expects a logarithmically slow
approach to the FP value of $\tilde \Gamma$. The only other possibility
is that $\gamma$ vanishes as a power law at the MIT. For this case
$\tilde \Gamma$ also vanishes as a power law, and in a scaling description
(cf. below) $\tilde \Gamma$ is a conventional irrelevant variable.
While at present we cannot exclude this scenario, we consider it
unlikely because of the first term on the r.h.s. of
Eq.\ (\ref{eq:3.6d}) (or the second term on the r.h.s of
Eq.\ (\ref{eq:2.15e})),
which tends to drive $\gamma_{c}$ and hence $\gamma$ towards larger values.

\subsection{Logarithmic Corrections to Scaling}
\label{subsec:IV.B}

In the previous subsection we have argued that in general
one expects $\tilde \Gamma$
to approach its fixed point value logarithmically slowly. This result is
consistent with
Eq.\ (\ref{eq:3.6d}) which gives $\gamma_{c} \rightarrow \gamma_{c}^{*}$
at the MIT, which in turn implies that $\tilde \Gamma$ vanishes
logarithmically slowly at the MIT. Because $\tilde \Gamma$ couples to all
physical quantities, cf.
Eqs.\ (\ref{eqs:3.6}), we conclude that in all universality
classes where Cooperons are present, logarithmic corrections to scaling
will appear. Note that this conclusion is independent of the spatial
dimensionality and depends only on whether or not the kernel $\gamma$ has a
constant contribution at the MIT. We also note that this is a
zero-temperature, quantum mechanical effect that might be relevant for other
quantum phase transitions.

For a specific example of an observable quantity, let us consider the
electrical conductivity $\sigma$, which is related to the disorder by,
\begin{equation}
\sigma=8S_{D}b^{-\epsilon}/(2\pi)^{D}\pi g(b)\quad.
\label{eq:4.5}
\end{equation}
Note that in giving
Eq.\ (\ref{eq:4.5}) we have used units such that $e^{2}/\hbar$
is unity, and we have ignored the possibility of charge renormalization.
For a discussion of the latter point we refer the reader elsewhere.\cite{25}
With t the dimensionless distance from the critical point at zero
temperature, and $\delta \tilde \Gamma=\tilde \Gamma - \tilde \Gamma^{*}$,
the conductivity satisfies the scaling equation,
\begin{equation}
\sigma(t,T)=b^{-\epsilon}\,{\cal F}[b^{1/\nu}t,b^{z}T,%
     \delta \tilde \Gamma(b)]\quad.
\label{eq:4.6}
\end{equation}
Here $\nu$ is the correlation length exponent, z is the dynamical scaling
exponent, and of the irrelevant variables in the scaling function $\cal F$
we have kept only the one that decays most slowly at the MIT, i.e.
$\delta \tilde \Gamma$.

At zero temperature we let $b=t^{-\nu}$, and assume that ${\cal F} [1,0,x]$ is
an analytic function of x since it is evaluated far from the MIT. We obtain,
\begin{eqnarray}
\sigma (t\rightarrow 0,T=0) \approx \sigma_{0} t^{s} \biggl\{ 1+%
     {a_{1} \over \ln (1/t)} + {a_{2} \over [\ln (1/t)]^{2}}
\nonumber\\
 + \ldots \biggr\}\quad,
\label{eq:4.7}
\end{eqnarray}
with $s=\nu (D-2)$. Here $\sigma_{0}$ is an unknown amplitude, and the
$a_{i}$ are unknown expansion coefficients. Depending on what the
subleading behavior of $\delta\tilde\Gamma (b)$ is, the $a_{i}$ with
$i\ge 2$ could carry a very weak t-dependence (e.g. $a_{2}\sim \ln \ln t$).

An interesting consequence of the logarithmically marginal operator
$\tilde\Gamma$ is that the dynamical scaling exponent in
Eq.\ (\ref{eq:4.6}) is
ill-defined. To see this we use that z is normally defined by,
\begin{equation}
z=d+\kappa\quad,
\label{eq:4.8}
\end{equation}
with $\kappa$ the exponent that determines the anomalous dimension of h.
The Eq.\ (\ref{eq:3.6b}) suggests that in general
the linearized RG flow equation for h will have the form,
\begin{equation}
{d\ln\,h \over dx}= \tilde \kappa + \alpha \delta \tilde \Gamma%
     \quad,
\label{eq:4.9}
\end{equation}
with $\alpha$ a universal constant and $\tilde \kappa=\kappa$ if
$\delta \tilde \Gamma$ vanishes as a power law. However, if
$\delta \tilde \Gamma$ vanishes logarithmically at the MIT, then,
\begin{equation}
h(b)=b^{\tilde \kappa}(\ln\,b)^{\alpha}\quad,
\label{eq:4.10}
\end{equation}
i.e., h does not scale as $b^{\tilde \kappa}$. This in turn implies that
Eq.\ (\ref{eq:4.6}) should be replaced by,
\begin{equation}
\sigma(t,T)=b^{-\epsilon}\,{\cal F}[b^{1/\nu}t,b^{D}h(b)T,%
     \delta \tilde \Gamma(b)]\quad.
\label{eq:4.11}
\end{equation}
At $t=0$ and as $T\rightarrow 0$, the Eqs.\ (\ref{eq:4.10}) and
(\ref{eq:4.11}) give,
\begin{eqnarray}
\sigma(t=0,T\rightarrow 0) \approx \sigma_{0} T^{\epsilon/z}%
     [\ln(1/T)]^{\epsilon \alpha /z}
\nonumber\\
     \times \biggl\{1-{\epsilon \alpha^{2} \over z}
     {\ln \ln(1/T^{\epsilon /z})
     \over \ln (1/T)} + \ldots \biggr\}\quad.
\label{eq:4.12}
\end{eqnarray}
We conclude that for $\sigma(t=0,T)$ the asymptotic scaling is in general
determined by logarithms.

\section{EXPERIMENTAL RELEVANCE}
\label{sec:V}

\subsection{Experiments in Zero Magnetic Field}
\label{subsec:V.A}

It is well known that for the experimental determination of critical
exponents, and for the comparison of theoretical and experimental values for
these quantities, one must take into account corrections to scaling. This
is so mainly because the asymptotic critical region, where corrections are
negligible, is too small to be experimentally accessible. Furthermore, a
reliable determination of the critical exponents usually requires
experimental data that cover many decades of the control parameter.\cite{27}
For conventional phase transitions, where the control parameter is the
temperature which is relatively easy to control, these conditions can be
met. In the case of the MIT the situation is much less favorable. The main
reason is that changing the control parameter, i.e. the impurity
concentration, usually requires the preparation of a new sample. The only
known way to avoid this problem is the stress-tuning technique
of Ref.\ \onlinecite{6}.
Also, since the transition occurs at $T=0$, measurements at very low
temperatures and a careful extrapolation to $T=0$ are required.
The application of these techniques to Si:P has led to the most accurate
determination of the critical behavior of a MIT system to date.
\cite{6,28} Still,
by the standards of critical phenomena experiments the data taken are
relatively far from the critical point, with $t\ge 10^{-3}$, and
corrections to scaling have not been considered in the data analysis.
This means that the measured exponent for the conductivity cannot be
identified with the asymptotic critical exponent s. Rather, the measured
value must be taken to represent some effective exponent $s_{eff}$, which
is different from s due to corrections to scaling. This observation offers
an interesting possibility to explain the discrepancy between the measured
value 0.51 for s, and the theoretical bound $s\ge 2/3$:\cite{9} If the
measured value represents an effective exponent, then the latter is not
subject to the theoretical constraint for s. According to this hypothesis,
the discrepancy between the measured value $s_{eff}=0.51$ and the
theoretical lower bound $s\ge 0.67$ must then be due to the corrections
to scaling. With the usual, power-law corrections such a large discrepancy
would be hard to explain. In the present case, however, where the corrections
are logarithmic, it turns out that they provide a viable explanation for
the observations.

We can use Eq.\ (\ref{eq:4.7}) to reconcile experiments\cite{6,7,8}
near the MIT which seem to give $s<2/3$
with the theoretical bound \cite{9} $s\ge 2/3$. In
Fig.\ \ref{fig:1} we show experimental data, extrapolated to zero temperature,
for the conductivity of Si:P. The data points were chosen as follows.
For small t, roundoff due to sample inhomogeneities sets a limit at
$t\simeq 10^{-3}$.\cite{6,28} At large t, at some point one leaves the
region where scaling, even with corrections taken into account, is valid. We
chose to include points up to $t=10^{-1}$. Obviously, for $t\rightarrow 1$ the
concept of corrections to scaling loses its meaning, and the expansion,
Eq.\ (\ref{eq:4.7}),
in powers of 1/ln$t$ breaks down. We assumed a standard deviation
of a quarter of the symbol size for all points except for the one at the
smallest t, where we assumed it to be three times as large. In order to
improve the statistics with many correction terms taken into account, we
augmented the thirteen data points for $10^{-3}\le t\le 10^{-1}$ by another
twelve points obtained by linear interpolation between neighboring points.
If all $a_{i}$ are set equal to zero then
a best fit to the data yields $s=0.51\pm 0.05 < 2/3$.\cite{6}
We now assume, somewhat arbitrarily, $s=0.7$, and use
Eq.\ (\ref{eq:4.7}) with the $a_{i} \not= 0$. We have used a standard $\chi^2$
fitting routine with singular value decomposition \cite{29} to optimize the
values of the $a_{i}$. The dotted,
dashed, and full lines, respectively, in Fig.\ \ref{fig:1}
represent the best fits
obtained with one, two, and three logarithmic correction terms. These fits
are of significantly higher quality than a straight line fit optimizing s.
More than three correction terms did not lead to further improvements in
the fit quality. While the value of s was chosen arbitrarily, this
demonstrates that this experient is certainly consistent with a lower
bound $s\ge 2/3$ once the logarithmic corrections to scaling are taken
into account. We have also tried to optimize s by choosing different
values for s and comparing the quality of the resulting fits with a fixed
number of coefficients $a_{i}$ taken into account (letting s float together
with the $a_{i}$ proved unstable). The result was a very shallow minimum in
the fit quality around s=0.7. Relatively large ($\pm 0.15$) fluctuations in
the best value of s were observed when large t data points were successively
eliminated. From our fitting procedure for this experiment we estimate
$s=0.70^{+0.20}_{-0.03}$, where the lower bound is set by the theoretical
bound rather than by the fit.

\subsection{Experiments in a Magnetic Field}
\label{subsec:V.B}

While this success in fitting the Si:P data is encouraging, one might
object that the model invoked contains an infinite set of unknown parameters,
namely the $a_{i}$ in
Eq.\ ({eq:4.7}), and that therefore the quality of the fit
obtained is of no significance. Also, since the value of the asymptotic
exponent in 3-D is not known, no quantitative statements can be made.
It is therefore important to see if the model can predict any
qualitative features that are independent of the actual value of s, and
whether or not such features are observed.

There are obvious features predicted by our model which follow from the
qualitative magnetic field dependence of the Cooperon. The first one
is a strong magnetic field dependence of the effective exponent $s_{eff}$.
If the logarithmic corrections to scaling are caused by the Cooper
propagator, which acquires a mass in a magnetic field, then any nonzero
field must act to destroy the logarithms. It thus follows from the model
that Si:P, or any material, in a magnetic field must show a value of
$s_{eff}$ that is larger than 2/3. Any observation
of an $s_{eff}$ smaller than 2/3 in a magnetic field would rule out the
logarithmic corrections to scaling as the source of the anomalously small
value of $s_{eff}$.

The second feature concerns the temperature dependence of the conductivity.
It is well known that those materials which show an $s_{eff} < 0$ also
show a change of sign of the temperature derivative $d\sigma / dT$ of the
conductivity as the transition is approached.\cite{7,30} Within the RG
description of the MIT this phenomenon can be explained by a change of
sign of the $g^{2}$-term in the g-flow equation,
Eq.\ (\ref{eq:3.6a}) or (\ref{eq:3.8a}).
Since it is observed in Si:B, for which
Eq.\ (\ref{eq:3.8a}) is the relevant flow
equation, as well as in Si:P, the change of sign cannot be associated
with the scaling behavior of the triplet interaction constant, but must
be due to $\tilde \Gamma$. Since a magnetic field suppresses the Cooperon
channel, this feature should therefore disappear in a sufficiently strong
magnetic field.

Let us make these predictions somewhat more quantitative. The relevant
magnetic field scale $B_{x}$ is given by the magnetic length
$l_{B}=\hbar c/eB$ being equal to the correlation length
$\xi \simeq k_{F}t^{-\nu}$. Let us assume, as we did in connection with
Fig.\ \ref{fig:1}, that the boundary of the critical region is given by
$t\simeq 0.1$.
With $k_{F} \simeq 4\times 10^{6} cm^{-1}$ for Si:B or Si:P near the MIT,
and with $\nu \simeq 1$ we then obtain $B_{x} \simeq 1T$. The model thus
predicts that for magnetic fields exceeding about 1T the observed effective
conductivity exponent should be larger than 2/3, and the change of sign
of $d\sigma /dT$ observed in zero field should disappear.

Both of these features have been observed in the recent experiments
by Sarachik {\it et al.}.\cite{7,30} The Cooperon induced logarithmic
corrections to scaling thus provide a consistent explanation for several
observed features of doped Silicon which otherwise would be mysterious.

\section{CONCLUSION}
\label{sec:VI}

In this paper we have reconsidered the disordered electron problem in
the presence of Cooperons, previous treatments of which had led to
conflicting results.\cite{1,2,10} In particular we have analyzed the
technical differences between Refs.\
\onlinecite{1,2}, which renormalized the Cooper propagator,
and Ref.\ \onlinecite{10}, which renormalized the Cooper vertex function.
Let us summarize the results of this analysis.

(1) Crucial differences exist between the electron-electron interaction
in the particle-hole channel and the particle-particle or Cooper channel,
respectively. In the particle-hole channel the conservation laws for
particle number and spin enforce an essentially scalar structure of the
propagators and vertex functions. As a result, vertex functions are simply
scalar inverses of propagators, and standard renormalization procedures
applied to either object lead to identical results. In contrast, in
the Cooper channel there is no conservation law which would lead to an
analogous simplification, and vertex functions and propagators are related
by complicated matrix inversion procedures, cf. Sec.\ \ref{sec:II}. This leads
to fundamentally different structures of the singularities in the two objects,

which poses a problem for the renormalization, cf. Sec.\ \ref{subsec:III.A}.

(2) An attempt to reduce the Cooper propagator renormalization to a single
RG flow equation encounters severe structural problems which we have
discussed in Sec.\ \ref{subsec:III.B}. At the Gaussian
level, the renormalizaton constant needed displays a BCS-like singularity
at a finite scale. This leads to a renormalized vertex function which is
not finite, but shows the same singularity. If one ignores this problem
and proceeds to derive a flow equation for the Cooper propagator, the
singularity in the Gaussian renormalization constant produces imaginary
terms in the flow equation at one-loop level which resemble the renormalon
singularity known in quantum field theory. This problem is nonperturbative
in nature with respect to the Cooper interaction constant $\gamma_{c}$.
This explains why it went unnoticed in Refs.\ \onlinecite{1,2}, which
expanded in both
$\gamma_{c}$ and in the disorder g. It invalidates the simple fixed point
structure found in these references, according to which the Cooper channel
interaction is a conventional irrelevant operator.

(3) Any renormalization scheme for the Cooper vertex function must
acknowledge the fact that the object that appears in perturbation theory
is the Cooper propagator $\tilde \Gamma$, not just the Cooper interaction
amplitude $\gamma_{c}$. In fact the propagator plays the role of an
effective interaction amplitude, cf. Sec.\ \ref{subsec:III.A}.
Due to the inversion problem mentioned in point (1) above it is a much more
complicated object than $\gamma_{c}$. This point was missed in
Ref.\ \onlinecite{10}. The
procedure there was effectively equivalent to inserting the zero-loop order
result for $\tilde \Gamma$ into the one-loop order flow equations for the
other coupling constants. The RG flow equations of Ref.\ \onlinecite{10}
are therefore not consistent in general. They are valid near the
transition only if $\gamma_{c}$ has a finite FP value.

(4) A possible solution of these problems is to derive flow equations for
all coupling constants in terms of $\tilde \Gamma$, and to determine the
latter from an Eliashberg-type integral equation which deals with the
inversion problem. This is the approach taken in Sec.\ \ref{sec:IV}.
Solutions of
the integral equation obtained for separable model kernels suggest that
$\tilde \Gamma$ is not a simple scaling variable and does not satisfy
an autonomous differential equation with universal coefficients. Plausible
assumptions about the kernel of the integral equation lead to the conclusion
that logarithmic corrections to scaling, as predicted in
Ref.\ \onlinecite{10}, should
exist regardless of whether $\tilde \Gamma$ approaches a nonzero fixed
point value, as asserted in Refs.\ \onlinecite{1,2}, or vanishes
logarithmically at large scales, as assumed in Ref.\ \onlinecite{10}.
However, this conclusion could not be made
mathematically precise since the full integral equation has not been
solved.

(5) Logarithmic corrections to scaling, as discussed in
Sec.\ \ref{subsec:IV.B}, can
explain some otherwise mysterious properties of materials that belong to
universality classes with Cooperons, cf.
Sec.\ \ref{sec:V}. The discrepancy between the
asymptotic critical exponent for the conductivity and the observable
effective exponent in this case is large enough to provide an explanation
for the 'exponent puzzle' in Si:P. This explanation is bolstered by two
predictions which have been verified by recent experiments:\cite{30}
In the presence of a sufficiently strong magnetic field the observed
conductivity exponent s must satisfy the lower bound $s \ge 2/3$, and the
characteristic change in the temperature dependence of the conductivity
observed in zero field as the transition is approached should disappear.

While these results are encouraging, some serious problems remain, most
notably the question of whether or not the theory is renormalizable. We
recall (cf. the discussion in
Sec.\ \ref{subsec:III.A}) that the pure nonlinear sigma-model
is renormalizable with two renormalization constants. The proof of this
makes heavy use of the symmetry properties of the model.\cite{21} Since the
interaction terms in the action break the symmetry of the sigma-model,
the proof of Ref.\ \onlinecite{21} is no longer applicable for the
interacting model
and renormalizability has never been proven. It is conceivable, however, that
the conservation laws in the particle-hole channel are still sufficient
to ensure renormalizability for the model without Cooperons. This would
be consistent with the considerable body of perturbative
evidence of renormalizability for this case. For the case with Cooperons
the absence of a conservation law and the unusual structure encountered
in attempts to apply renormalization techniques makes one much more
doubtful. It would be useful to further study this question. For instance,
it would be interesting to see whether the conservation laws are indeed
sufficient to ensure renormalizability. It would also be useful to have
an alternative approach to the subject which avoids the generalized
nonlinear sigma-model with its many technical problems. One could, for
instance, try to work directly with the Grassmannian action as in
recent work on interacting fermions without disorder.\cite{31}
One could also imagine an order parameter description of the MIT, using
the Q-field theory, but not the sigma-model.
Work in these directions is underway and will be presented in future
publications.

\acknowledgments

This work was supported by the NSF under grant numbers DMR-92-17496 and
DMR-92-09879.

\begin{figure}
\caption{Zero-temperature conductivity of Si:P vs. electron concentration.
      The data points have been redrawn from Ref.\
      \protect\onlinecite{6}. The
      dotted, dashed, and full lines, respectively, are best fits using
      one, two, and three corrections terms in Eq.\
      (\protect\ref{eq:4.7}) with an
      asymptotic critical exponent $s=0.7$. Best fit values for the
      coefficients in Eq.\
      (\protect\ref{eq:4.7}) are $\sigma_{0} = 54.87 ; 94.82 ;
      132.16 ; a_{1} = -1.84 ; -4.42 ; -6.61 ; a_{2} = 0 ; 6.21 ; 17.73 ;
      a_{3} = 0 ; 0 ; -16.38$ for the three curves, respectively. See the
      text for more details.}
\label{fig:1}
\end{figure}


\begin{references}
\bibitem{1} A.~M. Finkel'stein, Z. Phys. B {\bf 56}, 189 (1984);
     Zh. Eksp. Teor. Fiz. {\bf 84}, 168 [Sov. Phys. JETP {\bf 57}, 97.
\bibitem{2} C. Castellani, C. DiCastro, G. Forgacs,
     and S. Sorella, Solid State
     Commun. {\bf 52}, 261 (1984).
\bibitem{3} E. Abrahams, P.~W. Anderson, D.~C. Licciardello, and
     T.~V. Ramakrishnan, Phys. Rev. Lett. {\bf 42}, 673 (1979); for a review,
     see, P.~A. Lee and T.~V. Ramakrishnan, Rev. Mod. Phys. {\bf 57},
     287 (1985).
\bibitem{4} G. Bergmann, Phys. Rep. {\bf 101}, 1 (1984).
\bibitem{5} B.~L. Altshuler and A.~G. Aronov, Solid State Commun. {\bf 30},
115
     (1979); B.~L. Altshuler, A.~G. Aronov, and P.~A. Lee, Phys. Rev. Lett.
     {\bf 44}, 1288 (1980).
\bibitem{6} T.~F. Rosenbaum, R.~F. Milligan, M.~A. Paalanen, G.~A. Thomas,
     R.~N.Bhatt, and W.Lin, Phys. Rev. B {\bf 27}, 7509 (1983).
\bibitem{7} P. Dai, Y. Zhang, and M.~P. Sarachik, Phys. Rev. B {\bf 45}, 3984
     (1992).
\bibitem{8} W.~N. Shafarman, D.~W. Koon, and T.~G. Castner, Phys. Rev. B {\bf
40},
     1216 (1989).
\bibitem{9} J. Chayes, L. Chayes, D.~S. Fisher,
     and T. Spencer, Phys. Rev. Lett.
     {\bf 57}, 2999 (1986). This paper proved $\nu \ge 2/D$ with $\nu$ the
     correlation length exponent. For a heuristic argument to this effect,
     see, A.~B. Harris, J. Phys. C {\bf 7}, 1671 (1974). In order to conclude
     that $s \ge 2/D$ one
     needs in addition Wegner's (Z.Phys.B {\bf 25}, 327 (1976)) scaling law
     $s=\nu (D-2)$which holds in all existing theories of the MIT.
\bibitem{10} T.~R. Kirkpatrick and D. Belitz, Phys. Rev. Lett. {\bf 70}, 974
     (1993).
\bibitem{11} F. Wegner, Z. Phys. B {\bf 35}, 207 (1979).
\bibitem{12} see, e.g., G. Grinstein, in {\it Fundamental Problems in
     Statistical Mechanics VI}, edited by E.~G.~D. Cohen (North Holland,
     Amsterdam 1985), p.147.
\bibitem{13} D. Belitz and T.~R. Kirkpatrick, Phys. Rev. B {\bf 46}, 8393
(1992).
\bibitem{14} M. Grilli and S. Sorella, Nucl. Phys. B {\bf 295 [FS 21]}, 422
     (1988); D. Belitz and T.~R. Kirk\-pat\-rick, Nucl. Phys. B {\bf 316},
     509 (1989).
\bibitem{15}K.~B. Efetov, A.~I. Larkin,
     and D.~E. Khmelnitskii, Zh. Eksp. Teor.
     Fiz. {\bf 79}, 1120 (1980) [Sov. Phys. JETP {\bf 52}, 568].
\bibitem{16} T.~R. Kirkpatrick and D. Belitz, Phys. Rev. B {\bf 41},
     11082 (1990).
\bibitem{17} C. Castellani, C. DiCastro, P.~A. Lee, M. Ma, S. Sorella, and
     E. Tabet,
     Phys. Rev. B {\bf 30}, 1596 (1986); C. Castellani and C. DiCastro, Phys.
     Rev. B {\bf 34}, 5935 (1986); C. Castellani, C. DiCastro, G. Kotliar,
     P.A. Lee, and G. Strinati, Phys. Rev. Lett. {\bf 59}, 477 (1987);
     Phys. Rev. B {\bf 37}, 9046 (1988).
\bibitem{18} see, e.g., J. Zinn-Justin, {\it Quantum Field Theory and
Critical
     Phenomena} (Clarendon, Oxford 1989); M. Le Bellac, {\it Quantum and
     Statistical Field Theory} (Clarendon, Oxford 1991).
\bibitem{19} K.~G. Wilson  and J. Kogut, Phys. Rep. {\bf 12}, 75 (1974).
\bibitem{20} J. Polchinski, Nucl. Phys. B {\bf 231}, 269 (1984).
\bibitem{21} E. Br{\'{e}}zin,
     J. Zinn-Justin, and J.~C. Le Guillou, Phys. Rev. D
     {\bf 14}, 2615 (1976).
\bibitem{22} D.~R. Nelson and R.~A. Pelcovits, Phys. Rev. B {\bf 16},
     2191 (1977).
\bibitem{23} C. Castellani, C. DiCastro,
     and G. Forgacs, Phys. Rev. B {\bf 30},
     1593 (1984); C. Castellani, C. DiCastro, and S. Sorella, Phys. Rev. B
     {\bf 34}, 1349 (1986).
\bibitem{24} A.~M. Finkel'stein, in {\it Anderson Localization}, ed. by
     T. Ando and H. Fuku\-ya\-ma (Sprin\-ger, New York 1988).
\bibitem{25} D. Belitz and T.~R. Kirkpatrick, submitted to Phys. Rev. E.
\bibitem{26} T.~R. Kirkpatrick and D. Belitz, J. Phys. Cond. Matter {\bf 4},
L37 (1992).
\bibitem{footnote1} Often the RG is thought of as strictly a technique to
     resum logarithmic divergencies. If one adopts this interpretation one
     is tempted to expand Eq.\ \ref{eq:3.10a} in powers of $\gamma_{c} x$
     and ignore the singularity implied by the infinite series. We will
     show below Eq.\ \ref{eq:3.14} that such an expansion is invalid.
\bibitem{27} J.~M.~H. Levelt Sengers and J.~V. Sengers, in {\it Perspectives
     in Statistical Physics}, edited by H.~J. Raveche (North Holland,
     Amsterdam 1981) p.241.
\bibitem{28} Very recently, the interpretation of the data in
     Ref.\ \onlinecite{6} has been
     questioned (H. Stupp, M. Hornung, M. Lakner, O. Madel,
     and H.v. L\"{o}hneysen,
     Karlsruhe preprint 1993). These authors assert that the rounding of the
     data ascribed to sample inhomogeneities in Ref.\ \onlinecite{6}
     instead signals a
     crossover, and that the effective exponent is 1.3. In that case,
     logarithmic corrections to scaling may still be present, but in the
     absence of a theoretical value for s in D=3 the experiment would not be
     a suitable test of their presence. In what follows we show that even if
     the value of Ref.\ \onlinecite{6} is the correct one, the presence
     of logarithmic
     corrections to scaling can solve the exponent puzzle.
\bibitem{29} W.~H. Press, B.~P. Flannery, S.~A. Teutolsky,
     and W.~T. Vetterling,
     {\it Numerical Recipes} (Un\-iversity Press, Cambridge 1989),
     ch.14.
\bibitem{30} P. Dai, Y. Zhang, S. Bogdanovich, and M.~P. Sarachik, 'Critical
     Conductivity Exponent of Si:P in a Magnetic Field',
     City College preprint (1993).
\bibitem{31} G. Benfatto and G. Gallavotti,
     Phys. Rev. B {\bf 42}, 9967 (1990);
     J. Feldman and E. Trubowitz, Helv. Phys. Acta {\bf 63}, 157 (1990);
     {\bf 64}, 213 (1991);
     R. Shankar, 'Renormalization Group Approach to Interacting
     Fermions', Yale University preprint (1993).
\end{references}
\end{document}